\documentclass[aps,showpacs,preprint,superscriptaddress]{revtex4}
\usepackage{graphicx}
\usepackage{subfigure}
\usepackage{float}
\usepackage{amsmath}
\usepackage{amsfonts}
\usepackage{color}
\usepackage{bm}
\usepackage{bbm}
\usepackage{txfonts}
\usepackage{array}
\usepackage{amssymb}
\usepackage{multirow}
\makeatletter

\newcommand{\Rmnum}[1]{\expandafter\@slowromancap\romannumeral #1@}
\makeatother
\begin{document}

\title{Phase effect and symmetry on pair production in spatially inhomogeneous frequency chirping electric fields}
\author{Melike Mohamedsedik}
\affiliation{Key Laboratory of Beam Technology of the Ministry of Education, and College of Nuclear Science and Technology, Beijing Normal University, Beijing 100875, China}
\author{Lie-Juan Li}
\affiliation{Key Laboratory of Beam Technology of the Ministry of Education, and College of Nuclear Science and Technology, Beijing Normal University, Beijing 100875, China}
\author{Li Wang}
\affiliation{Institute of Radiation Technology, Beijing Academy of Science and Technology, Beijing 100875, China}
\author{Orkash Amat}
\affiliation{Key Laboratory of Beam Technology of the Ministry of Education, and College of Nuclear Science and Technology, Beijing Normal University, Beijing 100875, China}
\author{Li-Na Hu}
\affiliation{Key Laboratory of Beam Technology of the Ministry of Education, and College of Nuclear Science and Technology, Beijing Normal University, Beijing 100875, China}
\author{B. S. Xie \footnote{bsxie@bnu.edu.cn}}
\affiliation{Key Laboratory of Beam Technology of the Ministry of Education, and College of Nuclear Science and Technology, Beijing Normal University, Beijing 100875, China}
\affiliation{Institute of Radiation Technology, Beijing Academy of Science and Technology, Beijing 100875, China}
\date{\today}
\begin{abstract}
Effect of the carrier envelop phase on the electron-positron pair production is studied in spatially inhomogeneous electric field with symmetrical frequency chirping. In high or low original frequency field without chirping as well as one with chirping, we find that the strength of interference effect of the momentum spectrum and the reduced particle number are all changeable periodically with phase, in particular, these periodical changes are more sensitive to the applied parameters in case of low frequency field. At the small spatial scale, the reduced particle number change is over one order magnitude by phase in small chirping. For the reduced particle number, the different optimal phases are obtained at different spatial scales, however, the larger the chirping is applied, the higher the created pair number is got. Interestingly, some different types of symmetries, i.e., the mutual symmetry of mirror/coincidence for two correlated phases and the individual self symmetry for single phase, are unfolded on the momentum spectrum. The physical reason of the mutual symmetry between two correlated phases and also the individual symmetry for two fixed specific phases are examined and discussed analytically in detail. The combined roles by phase and chirping on the periodic and symmetrical behaviors of the momentum spectrum and the reduced particle number are expected to have the potential extension to more fields such as that with multidimensional spatial coordinate.
\end{abstract}
\maketitle

\section{Introduction}

A strong background field ensures the vacuum to produce real particles, for example electron-positron ($e^{-}e^{+}$) pair, that is a typical nonpurterbative prediction known as Schwinger effect in quantum electrodynamics (QED) \cite{Sauter:1931zz,Heisenberg:1935qt,Schwinger:1951nm}. It still waits an experimental verification~\cite{Gelis:2015kya} because
the electric field strength afforded by the current laser technology for pair production is rather smaller than the Schwinger critical field strength $E_{cr} =  {m_e^2c^3} / {e\hbar} \approx 1.3 \times 10^{18}  {\rm V}/{\rm m}$ (corresponding laser intensity~$I=4.3 \times 10^{29}{\rm W}/{\rm cm^{2}}$). However, the results of recent theoretical investigations suggest that vacuum $e^{-}e^{+}$ pair production is possible through careful combining and shaping of the laser pulse field even though the external field strength is one or two orders smaller than $E_{cr}$ \cite{Schutzhold:2008pz,Bell:2008zzb,DiPiazza:2009py,Bulanov:2010ei}. One may expect such schemes about the laser pulses will make experimental tests possible with the advanced laser facilities in the near future~\cite{Ringwald:2001ib,
Heinzl:2008an,Marklund:2008gj,Pike:2014wha}.

At about 50 years ago, vacuum pair production has been investigated in many original works \cite{Nikishov,Marinov,breyzin,Ritus,Nikishov1985} by using simple but very important fields such as constant electric and magnetic fields~\cite{Nikishov}, alternating electric field~\cite{Marinov,breyzin} and so on. Recent years, many theoretical analysis and numerical calculations have been performed based on previous researches, see Refs.~\cite{Hebenstreit:2009km,
Dumlu:2010vv,Hebenstreit:2011wk,Kohlfurst:2017hbd,Ababekri:2019dkl,Ababekri:2020,
Kohlfurst:2017git,Kohlfurst:2020,Akkermans:2011yn,Dumlu:2010ua,Dumlu:2011rr,Xie:2017,
Li:2015cea,Li:2017qwd,Olugh:2018seh,Wang:2019,Olugh:2019nej,Abdukerim:2013vsa,
Mohamedsedik:2021,Li:2021vjf,Wang}. The momentum spectrum and number density of created particles are examined analytically or/and numerically under both spatially homogeneous~\cite{Hebenstreit:2009km,Dumlu:2010vv,Akkermans:2011yn,
Dumlu:2010ua,Dumlu:2011rr,Xie:2017,Li:2015cea,Li:2017qwd,Olugh:2018seh,
Wang:2019,Olugh:2019nej,Abdukerim:2013vsa,Wang} as well as inhomogeneous electromagnetic fields~\cite{Hebenstreit:2011wk,Kohlfurst:2017hbd,Ababekri:2019dkl,Ababekri:2020,
Kohlfurst:2020,Mohamedsedik:2021,Li:2021vjf,Hu:2022} with many different field shapes. Recently, Fedotov \emph{et al.} have reviewed and summarized the key theories and progress in strong field QED in the past decade~\cite{Fedotov:2022}. These investigations enable us to understand new and challenging phenomena occurred in pair production process.

It is well known that the pair creation process is very sensitive to external field parameters such as field frequency, pulse duration, carrier envelop phase (CEP), pulse number, spatial scale and so on. The momentum spectrum or/and particle momentum distribution function display some different interesting structure even if these field parameters change small. In particular, pair production process has significant dependence on the CEP which is the phase difference between carrier wave and envelop of the pulse profile~\cite{Mackenroth,Banerjee}. Many investigations have been performed and the different nonlinear features of vacuum decay are observed for various oscillating forms of external fields \cite{Dumlu:2010vv,Abdukerim:2013vsa,Wang:2019,Banerjee,Aleksandrov,Bra,
Ababekri:2020,Mohamedsedik:2021}. Nevertheless, to our knowledge, most of them are merely concentrated on the role of CEP in time dependent background fields except Refs.~\cite{Ababekri:2020,Mohamedsedik:2021}, where only two typical cases of CEP have been considered with spatial field inhomogeneity. Therefore, it is necessary to study the CEP effect on pair production in spatially inhomogeneous electric fields systematically, and some important features of the momentum spectrum or/and reduced particle number with spatial inhomogeneity should be unfolded.

In this paper, we study the CEP effect on $e^{-}e^{+}$ pair production in spatially inhomogeneous fields with symmetrical frequency chirp. Field model is given as
\begin{equation}\label{FieldMode}
\begin{aligned}
E\left(x,t\right)
&=E_{0} f \left( x \right ) g\left( t \right )\\
&=\epsilon \, E_{cr} \exp \left(-\frac{x^{2}}{2 \lambda^{2}} \right ) \exp \left(-\frac{t^{2}}{2 \tau^{2}} \right ) \cos(b |t| t + \omega t + \varphi),
\end{aligned}
\end{equation}
where $\epsilon E_{cr}$ is the field strength, $\omega$ is the original center frequency, $\lambda$ and $\tau$ are the spatial and temporal scales, $b$ is symmetrical chirp parameter in the form $b=\alpha\omega/\tau$ ($\alpha \ge 0$)~\cite{Ababekri:2020,Mohamedsedik:2021}, and $\varphi$ is the CEP that in unit of $\varphi_{unit}=\pi/8$ for the numerical simplicity and convenience.

\begin{figure}[H]
\begin{center}
\includegraphics[width=\textwidth]{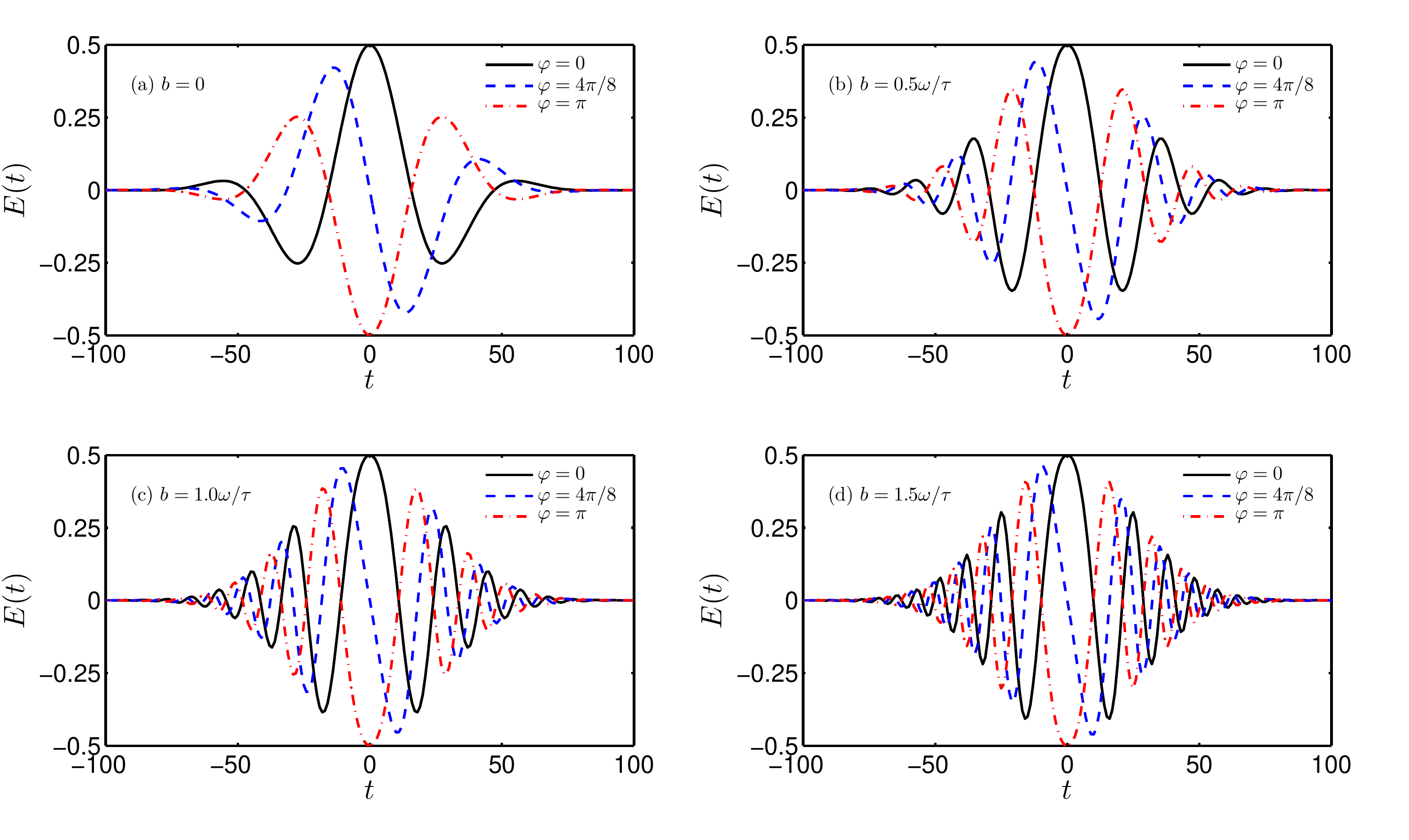}
\end{center}
\vspace{-12mm}
  \caption{(color online). Temporal electric field shape for $E\left(t\right) =0.5 E_{cr}\exp \left(-\frac{t^{2}}{2 \tau^{2}} \right ) \cos(b |t| t + \omega t + \varphi)$. The chirp values are $b=0$, $b=0.5\omega/\tau=0.002m^2$, $b=1.0\omega/\tau=0.004m^2$, $b=1.5\omega/\tau=0.006m^2$. Other field parameters are $\tau=25m^{-1}$, $\omega=0.1m$.}
  \label{fieldomg01}
\end{figure}

For a convenience to study the~CEP~effect in low frequency field, we plot the temporal electric field shape for various CEP with different chirp values in Fig.~\ref{fieldomg01}. We can see that the electric field changes from positive peaked electric field shape when $\varphi=0$ to two opposite signed peaked electric field when $\varphi=4\pi/8$, and it changes sign oppositely by the negative peaked electric field shape when $\varphi=\pi$. We note that the oscillation on the field shape also intensify with chirping. In the low frequency field, we will explain some interesting phenomena on the momentum spectrum by using this electric field shape.

In our study, we examine the significant influence of CEP on the momentum spectrum and the reduced particle number numerically by using the Dirac-Heisenberg-Wigner (DHW) formalism. We find that the interference effect of created particles reinforce with CEP. More importantly, in both of high and low- frequency fields, we have observed several different types of symmetries which named as exact mirror symmetry, approximated coincidence or/and mirror symmetry and approximated self symmetry, on the momentum spectrum for the different correlated translations of CEP, and analyzed by the field model as well as the equations of motion. The reduced particle number is enhanced periodically with increasing CEP and has symmetry in both of high and low-frequency electric fields. Moreover, we obtain the combined effect of CEP and symmetrical chirp on the momentum spectrum at different spatial scales. Meanwhile, we show the best optimal values of them which correspond to the maximum reduced particle number.

Note that the natural units ($\hbar=c=1$) are used throughout this paper and all quantities are expressed in terms of the electron mass $m$. Since the direction of the electric field is only one dimensional along the $x$-axis so that the study is limited to this direction by setting $p_{\perp}=0$.

\section{The DHW formalism}\label{method}

The DHW formalism is one of the quantum kinetic approaches developed from fermion density operator that is capable of describing many phase-space features of Dirac vacuum in QED~\cite{Bialynicki-Birula}. This formalism not only gives us the description of field-theoretic vacuum for $e^{-}e^{+}$ pair based on the phase-space concepts, but also enables us to perform the calculation within the standard formulation ~\cite{Bialynicki-Birula,Vasak:1987um} in both temporal~\cite{Olugh:2018seh,Olugh:2019nej,Wang} as well as spatial inhomogeneous electromagnetic fields~\cite{Hebenstreit:2011wk,Kohlfurst:2017hbd,Ababekri:2019dkl,Ababekri:2020,
Kohlfurst:2020,Mohamedsedik:2021,Li:2021vjf}. Since the details of the derivation of DHW formalism could be obtained in Refs.~\cite{Bialynicki-Birula,Vasak:1987um,Kohlfurst:2015zxi,Hebenstreit:2011pm}, we just give a brief presentation of DHW formalism  \cite{Kohlfurst:2015zxi,Hebenstreit:2011pm} for the sake of completeness of the paper.

Our study is focused on the pair production in time dependent spatially inhomogeneous electric field which is inhomogeneous on $x$-axis. So that the set of equations of motion
for Wigner components are reduced to four partial differential equations \cite{Kohlfurst:2015zxi} as
\begin{align}
 &D_t \mathbbm{s} - 2 p_x \mathbbm{p} = 0 , \label{pde:1}\\
 &D_t \mathbbm{v}_{0} + \partial _{x} \mathbbm{v}_{1} = 0 , \label{pde:2}\\
 &D_t \mathbbm{v}_{1} + \partial _{x} \mathbbm{v}_{0} = -2 m \mathbbm{p} , \label{pde:3}\\
 &D_t \mathbbm{p} + 2 p_x \mathbbm{s} = 2 m \mathbbm{v}_{1} , \label{pde:4}
\end{align}
with pseudodifferential operator
\begin{equation}\label{pseudoDiff}
 D_t = \partial_{t} + e \int_{-1/2}^{1/2} d \xi \,\,\, E_{x} \left( x + i \xi \partial_{p_{x}} \, , t \right) \partial_{p_{x}} .
\end{equation}
Here we define Wigner components $\mathbbm{w}_{0} = \mathbbm{s}$ , $\mathbbm{w}_{1} = \mathbbm{v}_{0}$ , $\mathbbm{w}_{2} = \mathbbm{v}_{1}$ and $\mathbbm{w}_{3} = \mathbbm{p}$. In order to perform the calculation, we employ the vacuum initial conditions \cite{Kohlfurst:2015zxi}
\begin{equation}\label{vacuum-initial}
{\mathbbm{w}_{0}}_{vac} = - \frac{2m}{\Omega} \, ,
\quad  {\mathbbm{w}}_{2 \, vac} = - \frac{2{ p_x} }{\Omega} \,  ,
\end{equation}
where $\Omega$ is the energy of single particle $\Omega=\sqrt{p_{x}^{2}+m^2}$. By subtracting initial vacuum terms in Eq.~(\ref{vacuum-initial}), we can obtain modified Wigner components $\mathbbm{w}_{k}^{v}$
\begin{equation}
\mathbbm{w}_{k}^{v} = \mathbbm{w}_{k}  - \mathbbm{w}_{vac},
\end{equation}
where $\mathbbm{w}_{k}$ is the Wigner component in Eqs. \eqref{pde:1}~-~\eqref{pde:4}.

The particle number density in phase space can be written as dividing the total particle energy with the individual particle energy\cite{Hebenstreit:2011wk}
\begin{equation}\label{particle number density}
n \left( x , p_{x} , t \right) = \frac{m  \mathbbm{s}^{v} \left( x , p_{x} , t \right) + p_{x}  \mathbbm{v}_{1}^{v} \left( x , p_{x} , t \right)}{\Omega \left( p_{x} \right)}.
\end{equation}
In this way, $n\left( p_x,t \right)$ or $n\left( x,t \right)$ could be obtained from $n \left( x , p_{x} , t \right)$ by integrating out $x$ or $p_x$, respectively.
The total number of created particles could be written
 \begin{equation}\label{Num}
N\left(t \right) = \int dx dp_x n \left( x , p_{x} , t \right).
\end{equation}
Moreover, in order to obtain the nontrivial spatial dependence of results on $\lambda$, we calculate the reduced quantities $n\left( p_{x}, t\rightarrow\infty \right)/\lambda \equiv\bar{n}\left( p_{x}, t\rightarrow\infty \right)$ and $N\left(t\rightarrow\infty\right)/{\lambda}\equiv\bar{N}\left(t\rightarrow\infty\right) \equiv \bar{N}$.

\section{CEP~effect on the momentum spectrum}\label{result1}

In this section, we study the CEP effect on the momentum spectrum in both high and low- frequency inhomogeneous fields with symmetrical frequency chirping. The momentum spectrum for various CEP at different spatial scales are shown and analyzed.

\subsection{High frequency field}\label{result11}
\begin{figure}[H]
\begin{center}
\includegraphics[width=\textwidth]{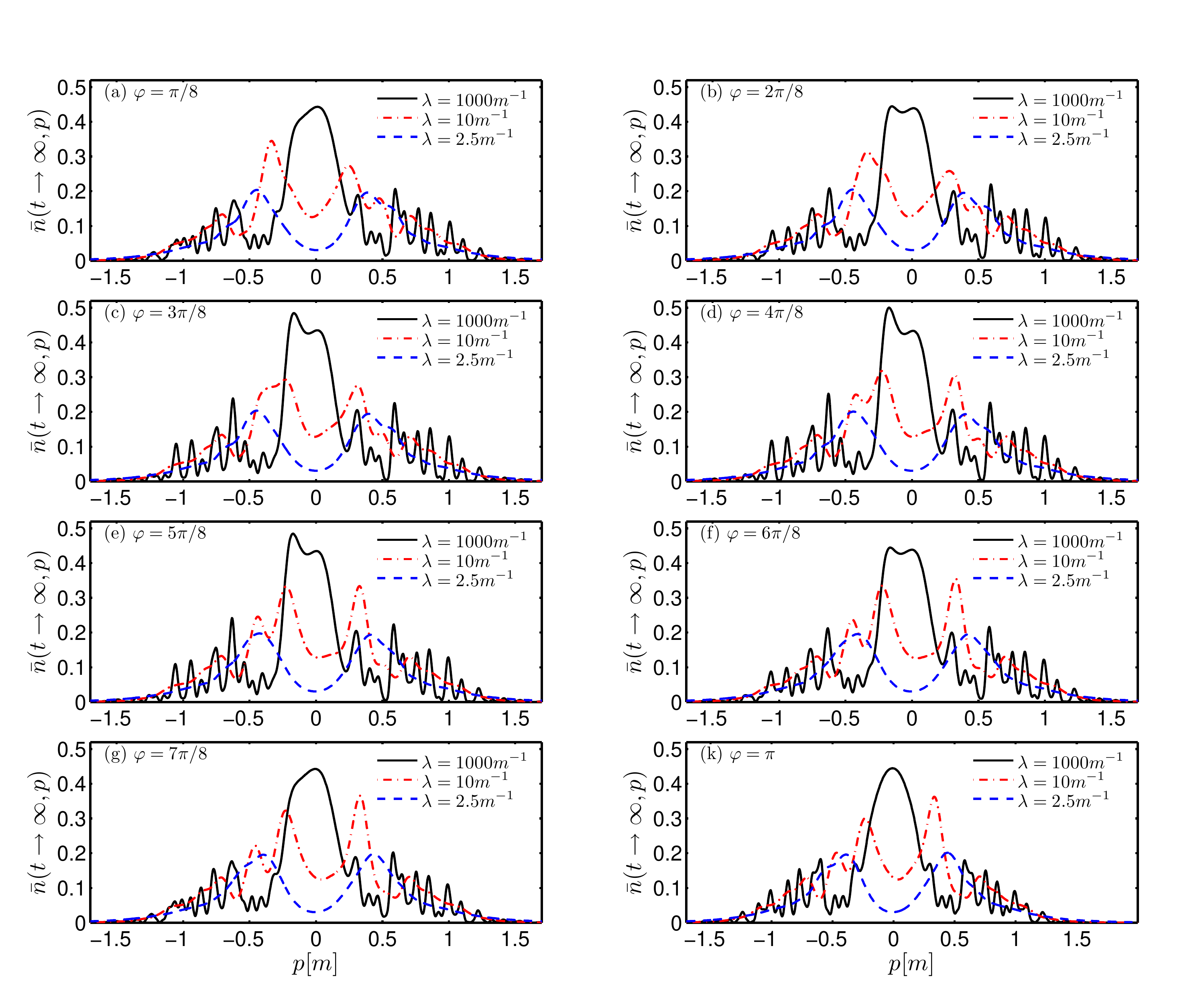}
\end{center}
\vspace{-12mm}
  \caption{(color online). The reduced momentum spectrum for various CEP in high frequency field at different spatial scales. The field parameters are $\epsilon=0.5$, $\omega=0.7 m$, $\tau=45m^{-1}$ and $b=0.2\omega/\tau \approx 0.0031 m^2$.}
  \label{omg07b02}
\end{figure}

In high frequency field, we set $\omega=0.7m$, $\tau=45m^{-1}$ and the pair is predominantly produced by multiphoton absorbtion. The CEP effect on the momentum spectrum at different spatial scales when $b=0.2 \omega/\tau \approx 0.0031 m^2$ are shown in Fig.~\ref{omg07b02}.
At large spatial scale, $\lambda=1000m^{-1}$, obvious oscillation is observed on the momentum spectrum, and it has periodically changed with increasing CEP. These oscillations are understood as interference effect of created particles by multiphoton process. The main peak corresponding to $p=0$ splits first and then recovers to the original shape with increase of CEP, see Fig.~\ref{omg07b02}. Note that the momentum peak amplitude is also periodically changeable with CEP.

When spatial scale decreases to $\lambda=10 m^{-1}$, we can see the weak oscillation on the momentum spectrum and it is almost unchangeable with CEP. However, the peak value on the left side of momentum spectrum decreases while the right peak increases gradually. Finally, the shape of left peak in Fig.~\ref{omg07b02}~(k)~becomes the right one in Fig.~$1$~(c)~of Ref.~\cite{Mohamedsedik:2021}~and vise versa, which reveals an exact mirror image symmetry holds for two correlated phases as $\varphi \leftrightarrow \pi+\varphi$, which will be discussed in the next section.

At very small spatial scale, $\lambda=2.5m^{-1}$, we still see two momentum peaks which are caused by the ponderomotive force in the cases of without/small chirping \cite{Kohlfurst:2017hbd,Ababekri:2020,Mohamedsedik:2021}. The weak dependence of the momentum spectrum on CEP can be attributed to a fact that the spatial scale and frequency chirping are too small to affect the role of the ponderomotive force that determines mainly the momentum spectrum.

\begin{table}[H]
\caption{Peak values of the reduced momentum spectrum for various CEP in high frequency field with different spatial scales when $b=0.2\omega/\tau \approx 0.0031 m^2$. Here $\bar{n}_{main}~(\lambda~[m^{-1}])$ is main peak value, $\bar{n}_{l}~(\lambda~ [m^{-1}])$ and $\bar{n}_{r}~(\lambda~[m^{-1}])$ are the peak values on the left and right sides of momentum spectrum, respectively. Other field parameters are $\epsilon=0.5$, $\omega=0.7m$ and $\tau=45m^{-1}$.}
\centering
\begin{ruledtabular}
\begin{tabular}{cccccc}
$\varphi$ &$\bar{n}_{l}(2.5)$&$\bar{n}_{r}(2.5)$ &$\bar{n}_{l}(10)$&$\bar{n}_{r}(10)$ &$\bar{n}_{main}(1000)$ \\
\hline
$0$         &$0.200$            &$0.196$             &$\mathbf{0.362}$&   $0.299$               & $0.445$\\
$\pi/8$     &$0.203$            &$0.196$             &$0.345$&            $0.274$               & $0.441$\\
$2\pi/8$    &$\mathbf{0.204}$   &$0.194$             &$0.313$&            $0.257$               & $0.445$\\
$3\pi/8$    &$\mathbf{0.204}$   &$0.194$             &$0.293$&            $0.276$               & $0.482$\\
$4\pi/8$    &$0.199$            &$0.193$             &$0.315$&            $0.305$               & $\mathbf{0.500}$\\
$5\pi/8$    &$0.197$            &$0.192$             &$0.333$&            $0.333$               & $0.484$\\
$6\pi/8$    &$0.195$            &$0.194$             &$0.334$&            $0.353$               & $0.445$\\
$7\pi/8$    &$0.196$            &$0.197$             &$0.321$&            $0.359$               & $0.442$\\
$\pi$       &$0.196$            &$\mathbf{0.200}$    &$0.299$&           $\mathbf{0.362}$       & $0.445$\\
\end{tabular}
\end{ruledtabular}
\label{Table 1}
\end{table}

In order to see the changes of peak values intuitively, we present the momentum peak values for various CEP at different spatial scales in Table~\ref{Table 1}. It is noticed that the maximum peak values mark with bold for the convenience reading in the table.

We can see that at large spatial scale, $\lambda=1000m^{-1}$, the largest peak value occurs for $\varphi=4\pi/8$. When $\lambda=10m^{-1}$, two peak values exchange places with each other for the left one at $\varphi=0$ ~\cite{Mohamedsedik:2021} and the right one at $\varphi=\pi$, just as we observed on the momentum spectrum. At very small spatial scale, $\lambda=2.5m^{-1}$, the peak value $\bar{n}_{l}$ decreases while $\bar{n}_{r}$ increases with CEP, finally two peaks exchange position with each other. However, the change on the peak value is very small for increasing CEP.
\begin{figure}[H]
\begin{center}
\includegraphics[width=\textwidth]{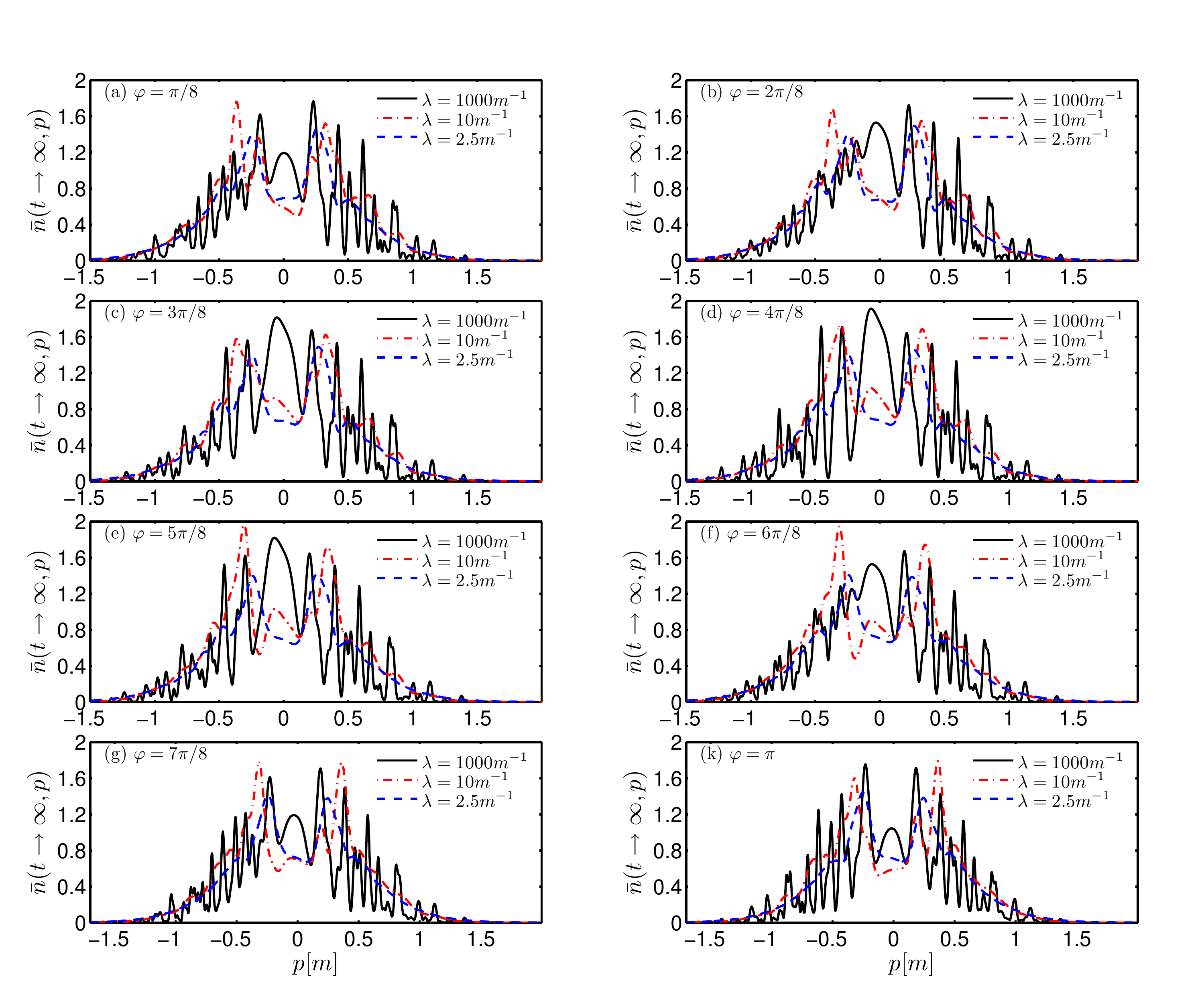}
\end{center}
\vspace{-15mm}
\caption{(color online). The reduced momentum spectrum for various CEP in high frequency field at different spatial scales when $b=0.5\omega/\tau \approx 0.0078 m^2$. Other field parameters are the same as in Fig.~\ref{omg07b02}.}
\label{omg07b05}
\end{figure}

Now let us to see the CEP effect on the momentum spectrum at different spatial scales when the frequency chirping becomes large as $b=0.5\omega/\tau \approx 0.0078 m^2$, which are shown in Fig.~\ref{omg07b05}.

At large spatial scale, $\lambda=1000m^{-1}$, a strong oscillation is observed on the momentum spectrum and it is more sensitive to CEP compared with the case in Fig.~\ref{omg07b02}. The different oscillation characteristic are the combined results of interference effect of created particles by the chirping \cite{Mohamedsedik:2021} and CEP. When large chirping is fixed, the oscillation is found intensifying with increasing CEP, and it reaches the maximal at $\varphi=4\pi/8$, however, it is weaken as increase of CEP when $\varphi>4\pi/8$. This combination of chirping and CEP results in the periodic variation of the interference effect between strong and weak. It implies that the momentum pattern can be also controlled by adjusting the CEP beside the frequency chirping.

At small spatial scales, $\lambda=10m^{-1}$ and $\lambda=2.5m^{-1}$, stronger oscillations also occur for increasing CEP compared to the case in Fig.~\ref{omg07b02}, especially when $\lambda=2.5m^{-1}$, see dashed blue line in Fig.~\ref{omg07b05}. It demonstrates that CEP plays a more remarkable role on momentum spectrum when chirp is large enough, where the CEP effect is suppressed greatly by the small spatial scale. The periodic changes of oscillation pattern and momentum peak values as well as the exchange phenomena of left- and right- peak are still observed the on momentum spectrum. For example, when $\lambda=10m^{-1}$, the peak value on the left side reaches $\bar{n}(t\rightarrow\infty, p)=1.936$ at $\varphi=5\pi/8$, see Fig.~\ref{omg07b05}(e), however, when $\varphi=6\pi/8$, it starts to decrease again.

It is noted that when $\varphi=\pi$, see Fig.~\ref{omg07b02}(k) and Fig.~\ref{omg07b05}(k), at all spatial scales, the momentum spectrum is just the inverse momentum spectrum of the case when $\varphi=0$, refer to Fig. 1 (c) and (d) in Ref.~\cite{Mohamedsedik:2021}. We find it as an exact symmetry of the mirror image momentum spectrum, which will be examined in the next section in detail.

\begin{table}[H]
\caption{Peak values of the reduced momentum spectrum for various CEP in high frequency field at different spatial scales when $b=0.5\omega/\tau \approx 0.0078 m^2$. Other field parameters are the same as Table~\ref{Table 1}. }
\centering
\begin{ruledtabular}
\begin{tabular}{cccccc}
$\varphi$ &$\bar{n}_{l}(2.5)$&$\bar{n}_{r}(2.5)$ &$\bar{n}_{l}(10)$&$\bar{n}_{r}(10)$ &$\bar{n}_{mean}(1000)$ \\
\hline
$0$         &$1.383$            &$1.425$             &$1.770$&            $1.583$                &$1.045$\\
$\pi/8$     &$1.386$            &$1.458$             &$1.749$&            $1.498$                &$1.194$\\
$2\pi/8$    &$1.375$            &$\mathbf{1.492}$    &$1.643$&            $1.545$                &$1.530$\\
$3\pi/8$    &$1.397$            &$1.485$             &$1.551$&            $1.623$                &$1.815$\\
$4\pi/8$    &$1.372$            &$1.441$             &$1.740$&            $1.672$                &$\mathbf{1.861}$\\
$5\pi/8$    &$1.388$            &$1.415$             &$\mathbf{1.936}$&   $1.710$                &$1.820$\\
$6\pi/8$    &$1.400$            &$1.387$             &$1.927$&            $1.722$                &$1.526$\\
$7\pi/8$    &$1.412$            &$1.372$             &$1.777$&            $1.761$                &$1.191$\\
$\pi$       &$\mathbf{1.425}$   &$1.383$             &$1.583$&            $\mathbf{1.770}$       &$1.045$\\
\end{tabular}
\end{ruledtabular}
\label{Table 2}
\end{table}
In Table~\ref{Table 2} we also present the momentum peak values for various CEP when $b=0.5\omega/\tau \approx 0.0078 m^2$. We can see that momentum peak values increase significantly in comparison with the case in Table~\ref{Table 1}, it is resulted by the combined effect of CEP and largest chirp. The periodical changes of the peak values and peak exchangeable in Table~\ref{Table 1} still exist. The changes of peak values at $\lambda=2.5m^{-1}$ is more obvious in comparison with that in the case of Table~\ref{Table 1}.

\subsection{Low frequency field}\label{result12}
\begin{figure}[H]
\begin{center}
\includegraphics[width=\textwidth]{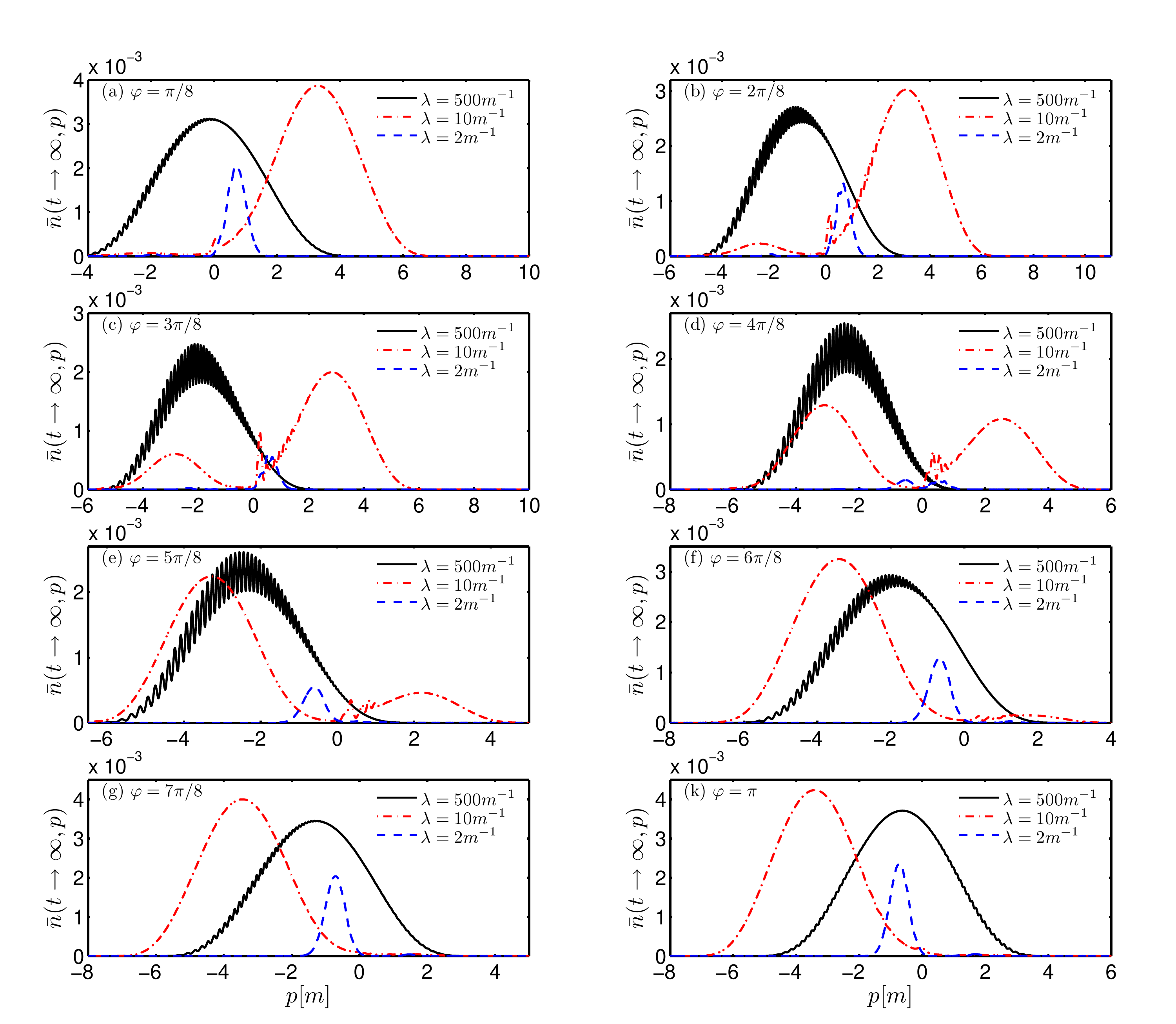}
\end{center}
\vspace{-12mm}
  \caption{(color online). The reduced momentum spectrum for various CEP in low frequency field at different spatial scales when $b=0$. Other field parameters are $\epsilon=0.5$, $\omega=0.1m$, $\tau=25m^{-1}$.}
  \label{omg01b0}
\end{figure}

In low frequency field, we choose $\omega=0.1m$ and $\tau=25m^{-1}$ and the pair is predominantly produced by tunneling process.
The CEP effect on momentum spectrum at different spatial scales when $b=0$ are shown in Fig.~\ref{omg01b0}.

At large spatial scale, $\lambda=500m^{-1}$, momentum spectrum shrinks first when $\pi/8 \le \varphi \le 3\pi/8$, then broadens and shifts towards negative direction when $4\pi/8\le \varphi \le \pi$. We think the reason can be explained by using the temporal electric field shape in Fig.~\ref{fieldomg01}~(a). The positive peaked electric field shape when $\varphi=0$ changes to two opposite signed electric field when $\varphi=4\pi/8$, and it changes sign oppositely by the negative peaked electric field shape when $\varphi=\pi$ finally. Such change behavior on the field shape induces the changes of motion of created particles in field region and it is reflected by momentum spectrum. We also note that the oscillations on the momentum spectrum are intensifying periodically with increasing~CEP. They can be understood as the interference effect of particles created by the electric field.
\begin{figure}[H]
\begin{center}
\includegraphics[width=\textwidth]{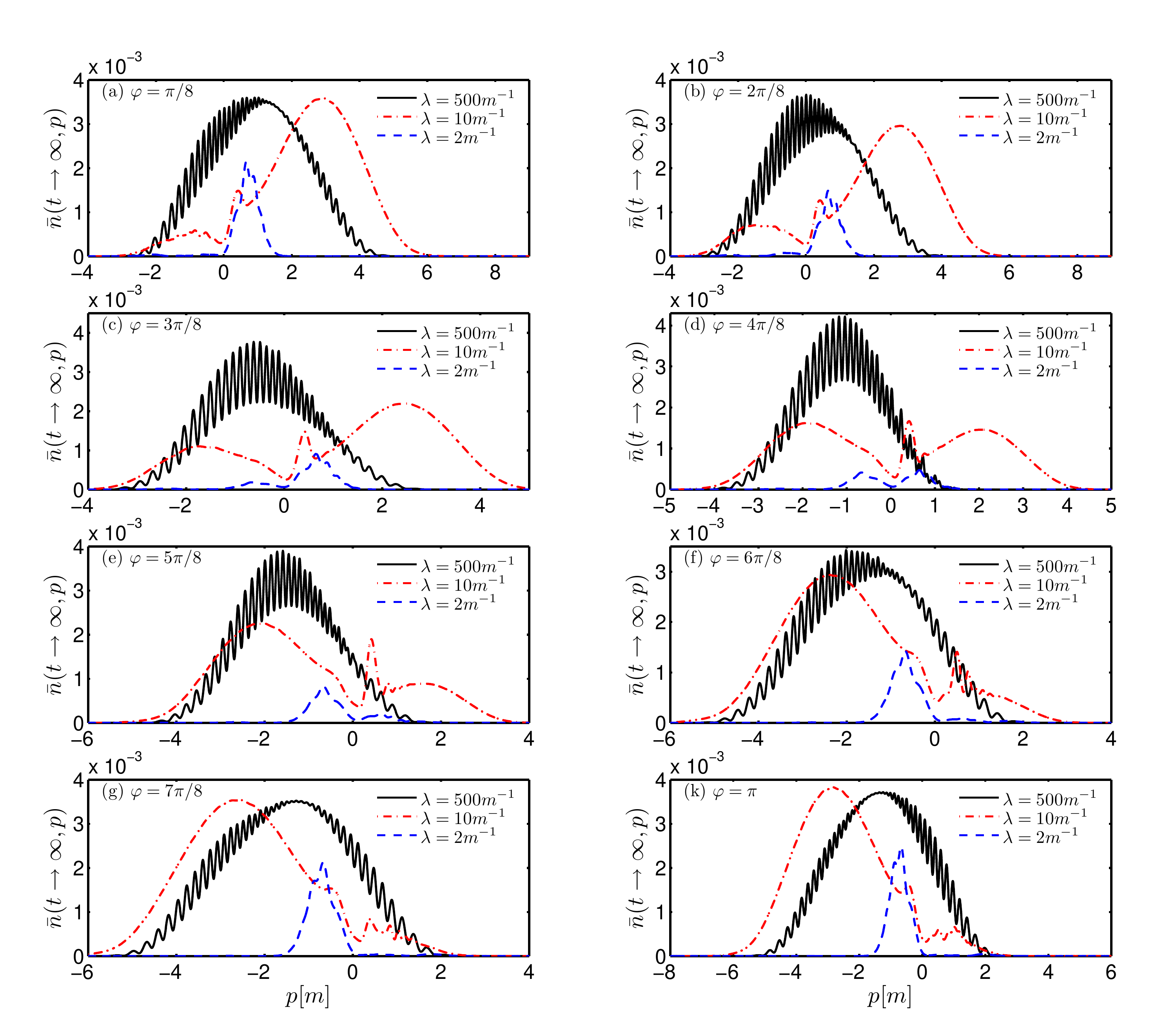}
\end{center}
\vspace{-12mm}
  \caption{(color online). The reduced momentum spectrum for various CEP parameters in the low frequency field at different spatial scales when $b=0.5\omega/\tau=0.002m^2$. Other field parameters are the same as in Fig.~\ref{omg01b0}.}
  \label{omg01b05}
\end{figure}
 Moreover, we can also find the periodical changes on the momentum peak values with CEP. For example, the peak values decrease for small CEP in Fig.~\ref{omg01b0}~(a)~-~(d), however, they increase again for large CEP in Fig.~\ref{omg01b0}~(e)~-~(k). These periodical changes on momentum spectrum are similar with that in the cases of high frequency field we have studied in the previous subsection.

At small spatial scales, $\lambda=10m^{-1}$ and $\lambda=2m^{-1}$, momentum spectrum also broadens and shifts towards negative field region with increasing CEP. It is because that the negative signed electric field is stronger and closer to negative peak value with increasing CEP. At the end, electric field changes sign and turns into negative peaked electric field when $\varphi=\pi$, see dashed-dotted red line in Fig.~\ref{fieldomg01}~(a). Therefore, the momentum spectrum also changes sign correspondingly, see Figs.~\ref{omg01b0}~(a)~-~(k) when $\lambda=10m^{-1}$ and $\lambda=2m^{-1}$, respectively (dashed-dotted red and dashed blue lines).
\begin{figure}[H]
\begin{center}
\includegraphics[width=\textwidth]{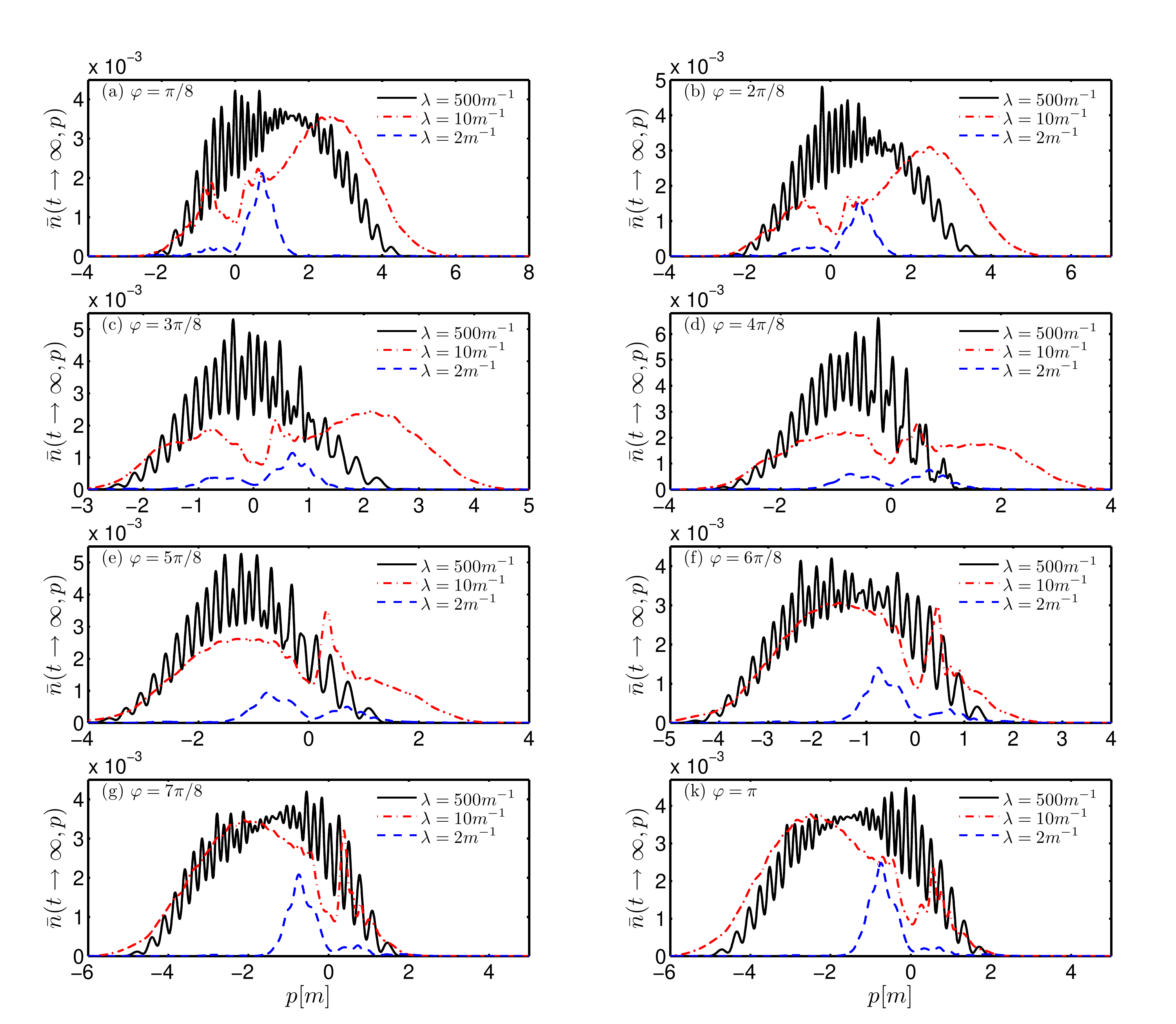}
\end{center}
\vspace{-12mm}
  \caption{(color online). The reduced momentum spectrum for various CEP in the low frequency field at different spatial scales when $b=1.0\omega/\tau=0.004m^2$. Other field parameters are the same as in Fig.~\ref{omg01b0}.}
  \label{omg01b1}
\end{figure}
Furthermore, the momentum peak value decrease gradually with increasing CEP until $\varphi=4\pi/8$. The two opposite positioned peaks (left and right) are observed when $\varphi=4\pi/8$. The reason is also that the field has two opposite peaks when $\varphi=4\pi/8$, see dashed blue line in Fig.~\ref{fieldomg01}(a), therefore, it leads to two opposite positioned momentum peaks correspondingly. We will give further discussion on this problem in Section.~\ref{discussion}.

\begin{figure}[H]
\begin{center}
\includegraphics[width=\textwidth]{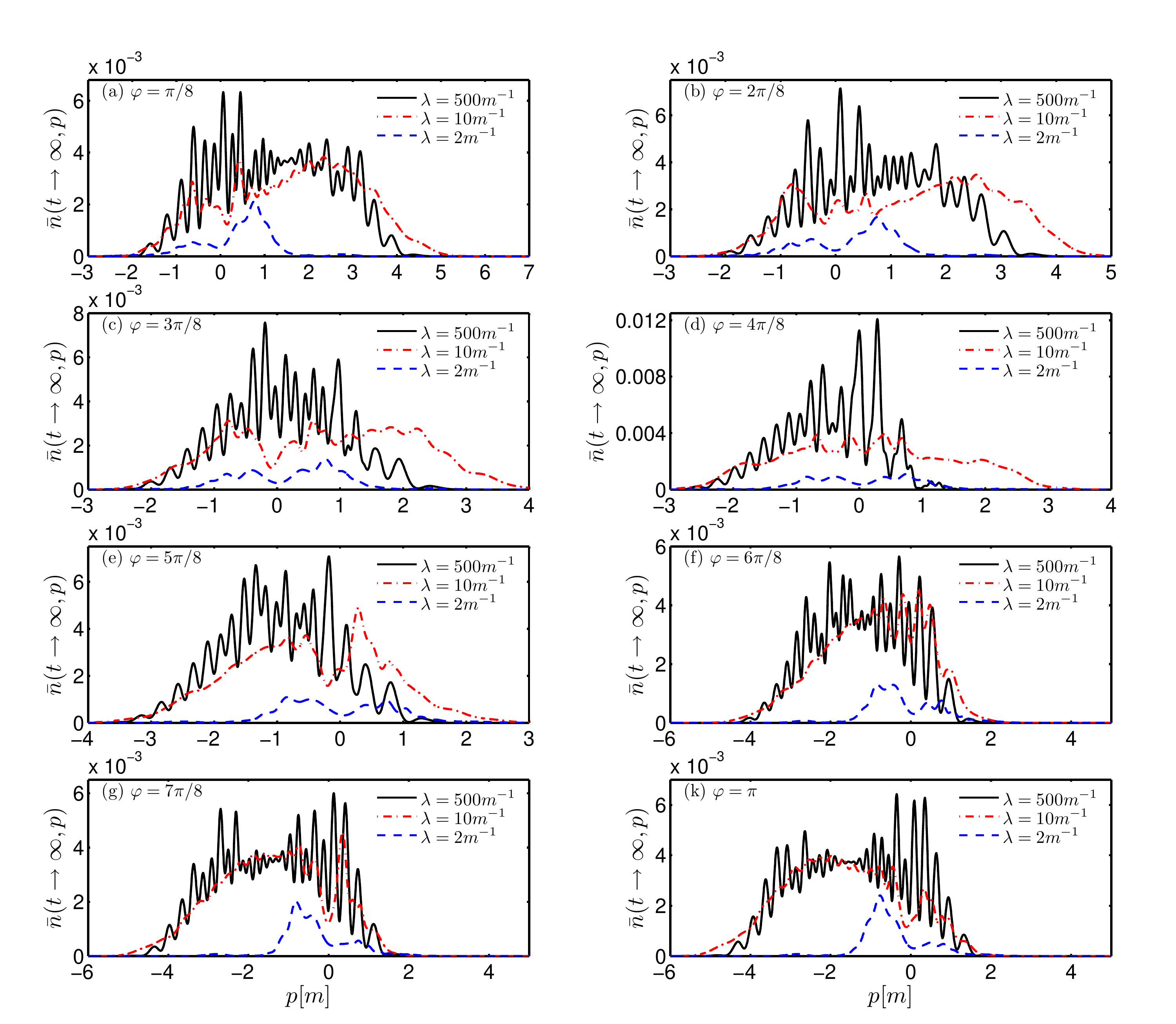}
\end{center}
\vspace{-12mm}
  \caption{(color online). The reduced momentum spectrum for various CEP in the low frequency field with different spatial scales when $b=1.5\omega/\tau=0.006m^2$. Other field parameters are the same as in Fig.~\ref{omg01b0}.}
  \label{omg01b15}
\end{figure}

In Figs.~\ref{omg01b05}~-~\ref{omg01b15}, we also plot momentum spectrum for various CEP at different spatial scales with chirp values $b=0.5\omega/\tau=0.002m^2$, $b=1.0\omega/\tau=0.004m^2$~and $b=1.5\omega/\tau=0.006m^2$, respectively.
At large spatial scale, $\lambda=500m^{-1}$, the periodically strengthened oscillations with increasing~CEP~still exist and are more significant for the largest chirp, see solid black lines in Figs.~\ref{omg01b05}~-~\ref{omg01b15}. These oscillations are also observed even at small spatial scale, $\lambda=10m^{-1}$, when chirp is large enough, also see dashed-dotted red lines in Figs.~\ref{omg01b05}~-~\ref{omg01b15}. For example, the opposite positioned peaks are still observed for small chirping case when~$\varphi=4\pi/8$~and they are replaced by strong oscillations for the largest chirp, see Fig.~\ref{omg01b05}(d)~-~Fig.~\ref{omg01b15}(d). These oscillations are caused by the different electric field shapes with various~CEP~which oscillates faster and faster with increasing chirp, see Fig.~\ref{fieldomg01}~(b)~-~(d), and it can also be understood as the interference effect of created particles. Again it is caused by the combined effect of CEP and chirping as the enforced interference of created particles in the view of Wentzel-Kramers-Brillouin (WKB) and turning points structure \cite{Mohamedsedik:2021}. We also note that at the very small spatial scale, $\lambda=2m^{-1}$, we could not find obvious oscillation on the momentum spectrum with increasing~CEP~even if the chirp becomes the largest, see the dashed blue lines in Figs.~\ref{omg01b05}~-~\ref{omg01b15}. Because that the field exists in very small spatial scale, the field energy is also very small correspondingly. By the way, the work done by the electric field would be reduced significantly. Consequently, very few particles are created from tunneling and unable to lead obvious interference effect on the momentum spectrum.

At all spatial scales, the momentum peak values also change periodically with~CEP, meanwhile increase significantly with increasing chirp. For example, the maximum peak value~$\bar{n}_{max}(500m^{-1})=0.012$~is observed when~$\varphi=4\pi/8$~for the largest chirp~$b=1.5\omega/\tau=0.006m^2$. Moreover, it is similar to that of high frequency field, we obtain exact mirror image spectrum in the low frequency field. For example, one can see the symmetry between Fig. 5 of Ref.~\cite{Mohamedsedik:2021} when $\varphi=0$ and Fig.~\ref{omg01b0}(k)~-~Fig.~\ref{omg01b15}(k) when $\varphi=\pi$ for four different chirping. We will give detailed explanation in the next section about this symmetry.

\section{the Symmetry on momentum spectrum}\label{symmetry}
In both of high and low frequency fields, we have observed various types of symmetries on the momentum spectrum which depend on CEP translation and spatial scale. Three different symmetries, named exact mirror, approximated coincidence or/and mirror, and approximated self symmetry, are found, identified and analyzed by the field model as well as the Eqs.~\eqref{pde:1}~-~\eqref{pde:4} of Wigner components. They correspond to the following cases: the exact mirror symmetry is for $\varphi\leftrightarrow\pi+\varphi$, which is associated the phases between the first and third quadrant, or the second and fourth one; the approximated coincidence and/or mirror symmetry are for $\varphi\leftrightarrow\pi-\varphi$, which is associated the phases between the first and second quadrant at large or/and small spatial
scale; they occur also for $\varphi\leftrightarrow2\pi-\varphi$, which is associated the phases between the first and fourth quadrant, or the second and third one at both of large and small spatial scales, respectively; Finally the approximated self symmetry occurs for the special cases of phase $\varphi=4\pi/8$ and $\varphi=12\pi/8$, where its shifted phase is $12\pi/8=\pi+ 4\pi/8=2\pi-4\pi/8$, therefore, it is the combination of exact mirror and approximated coincidence symmetries.

\begin{figure}[H]
\begin{center}
\includegraphics[width=\textwidth]{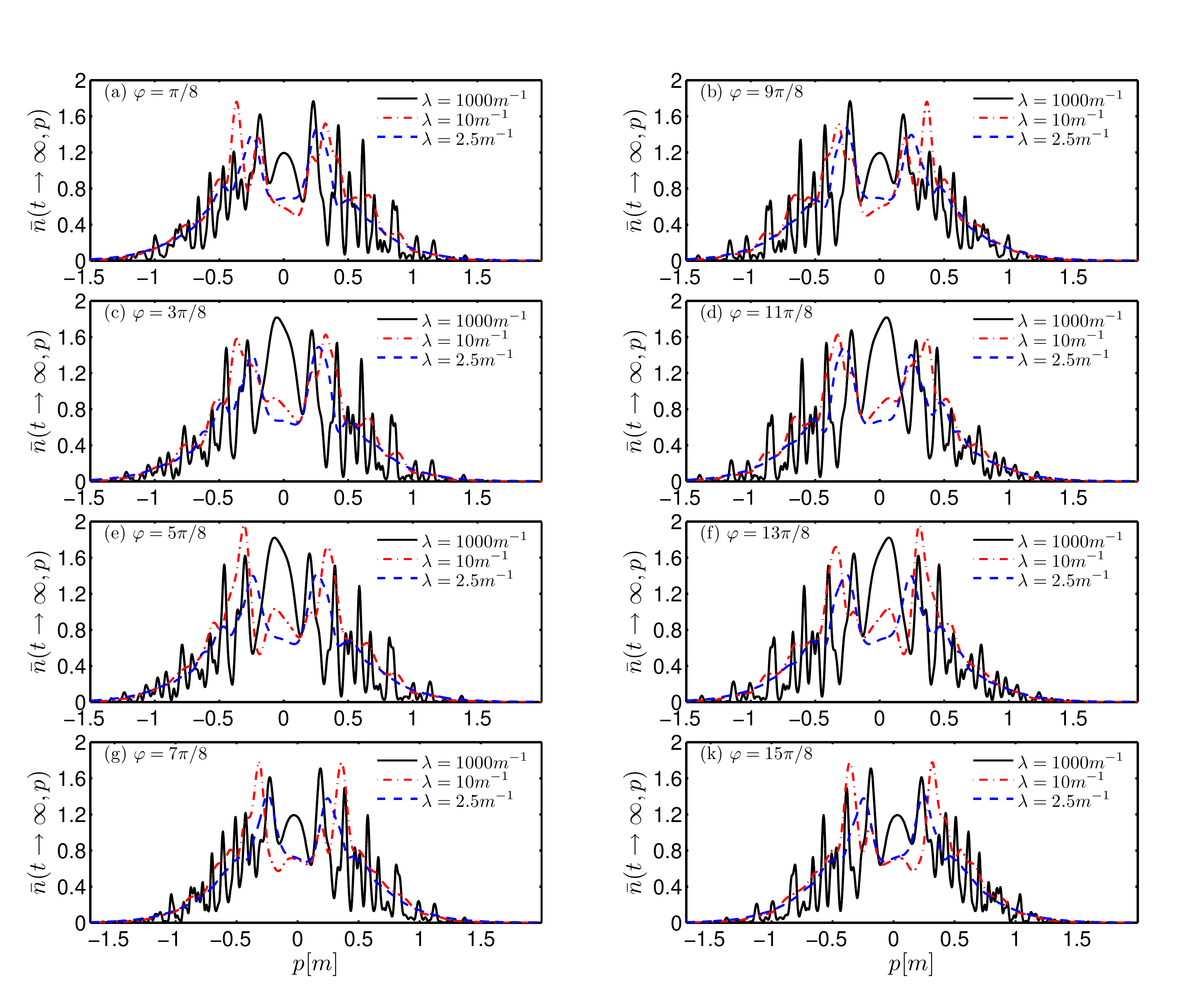}
\end{center}
\vspace{-12mm}
  \caption{(color online). The original (left) and exact mirror image (right) momentum spectrum for $\varphi$ and $\pi+\varphi$ at different spatial scales in high frequency field. Other field parameters are $\epsilon=0.5$, $\omega=0.7 m$, $\tau=45m^{-1}$ and $b=0.5\omega/\tau=0.0078m^2$.}
  \label{omg07mirror}
\end{figure}

In order to make a full understanding of these symmetries, we give a detailed analysis in Appendix~\ref{Appendix} for the field model and involving physical quantities for different phase relations.

\subsubsection{Exact mirror symmetry}

\begin{figure}[H]
\begin{center}
\includegraphics[width=\textwidth]{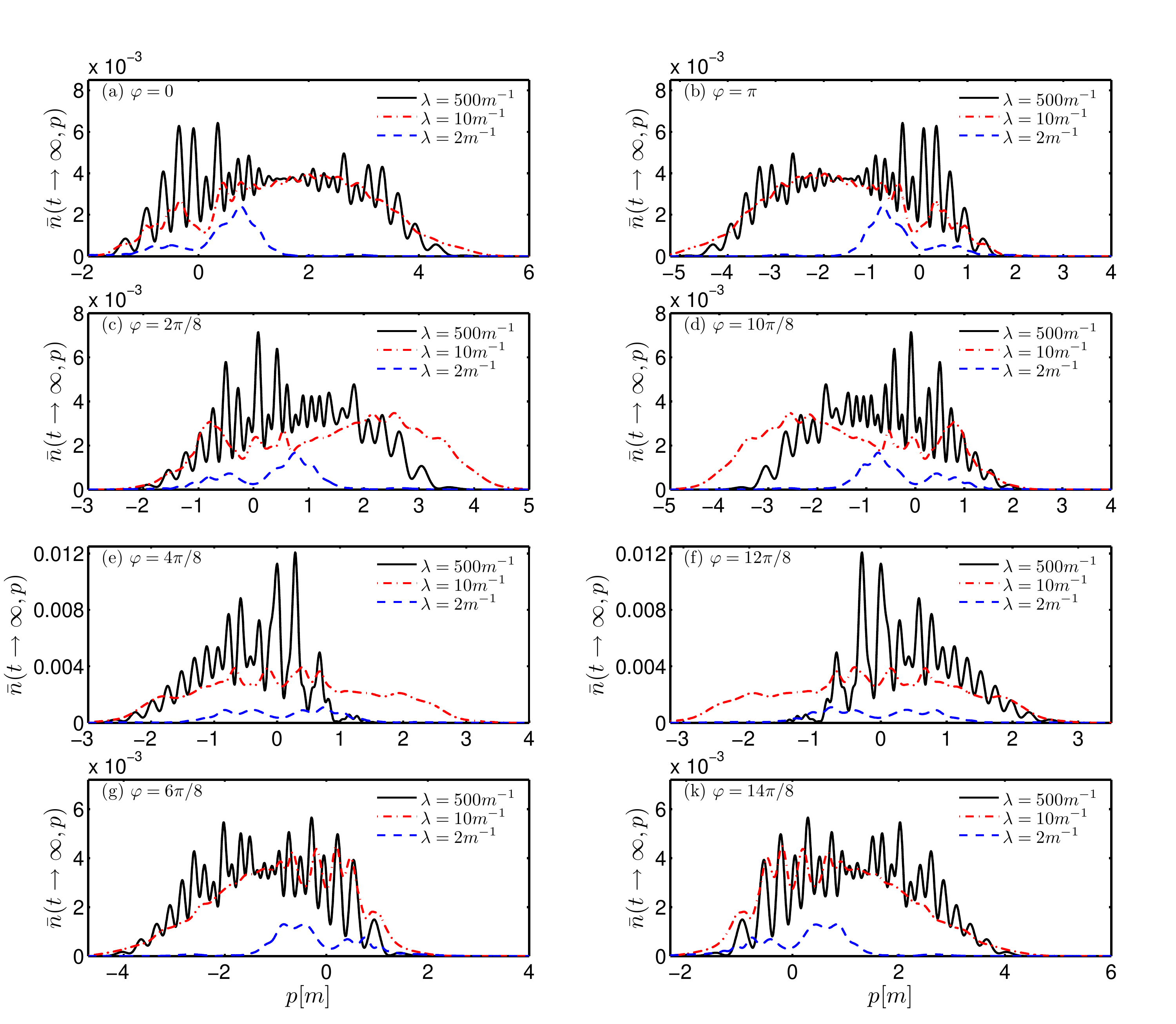}
\end{center}
\vspace{-12mm}
  \caption{(color online). The original and exact mirror image momentum spectrum for $\varphi$ (left) and $\pi+\varphi$ (right) at different spatial scales in low frequency field. Other field parameters are $\epsilon=0.5$, $\omega=0.1 m$, $\tau=25m^{-1}$ and $b=1.5\omega/\tau=0.006m^2$.}
  \label{omg01mirror}
\end{figure}

When $\varphi$ becomes $\pi+\varphi$, the momentum spectrum of them is just the exactly mirror image with each other so that we call it as the exact mirror symmetry. In order to see it intuitively, we have shown the momentum spectrum as the original and exact mirror one for $\varphi$ and $\pi+\varphi$ at different spatial scales in both of high and low- frequency fields, see Fig.~\ref{omg07mirror} and Fig.~\ref{omg01mirror}, respectively, when the fixed chirp is chosen the largest.
From two figures, we find that for each row of momentum spectrum, the spectrum on the right is the exact mirror image one of which on the left and vise versa. We also find that the exact mirror symmetry is strictly valid for all spatial scales as well as all chirp values. When CEP $\varphi$ shifts towards to $\pi+\varphi$, the electric field changes sign and propagates in opposite direction, see Eq.~(\ref{FieldModepi}). The created particles also move to opposite direction with opposite signed momentum in the field region and present such exact mirror image momentum spectrum as shown in Fig.~\ref{omg07mirror} and Fig.~\ref{omg01mirror}. We have checked and ascertained the signs of all physical quantities and Wigner components which correspond to exact mirror image spectrum. The results are summarized in Table~\ref{Table 3}, see the Case~\textbf{\Rmnum{1}} in Appendix~\ref{Appendix}. We find that, by a mirror transformation of momentum, all these physical quantities and Wigner components still satisfy equations of motion in Eqs. \eqref{pde:1}~-~\eqref{pde:4}. In this sense, we have proved the validity of exact mirror symmetry on momentum spectrum, which can explain or/and understand our numerical results in the previous section.

By the way, the periodical changes of the peak values and the peak exchangeable on the momentum spectrum which we mentioned in Table~\ref{Table 1} and Table~\ref{Table 2} can be also understood by the exact mirror symmetry on the momentum spectrum.

\subsubsection{Approximated coincidence or/and mirror symmetry}

For the following two cases, $\varphi\leftrightarrow\pi-\varphi$ and $\varphi\leftrightarrow2\pi-\varphi$, we have observed the approximated coincidence or/and approximated mirror symmetry on the momentum spectrum. It is worthy to point out that in these two cases, the symmetry is not strictly valid as in the first case. On the other hand, the types of approximated symmetries are different for large and small spatial scales. These different types of approximated symmetries are obtained in Appendix~\ref{Appendix} for two different correlated translations of CEP by the different approximate treating of the electric field at large and small spatial scales, respectively.

($1$) The approximated symmetry between $\varphi$ and $\pi-\varphi$.

When the addition of two correlated CEPs is $\pi$, i.e., $\varphi\leftrightarrow\pi-\varphi$: At large spatial scale, we observe the approximated coincidence symmetry on the momentum spectrum for three set of correlated CEPs as $\pi/8\leftrightarrow7\pi/8$, $2\pi/8\leftrightarrow6\pi/8$, $3\pi/8\leftrightarrow5\pi/8$, in both high and low frequency fields, see solid black lines in Figs.~\ref{omg07b02}~-~\ref{omg01b15}.
At small spatial scales, however, we observe the approximated mirror symmetry on the momentum spectrum for the three set of correlated CEPs that we just mentioned above, see dashed blue and dashed-dotted red lines in Figs.~\ref{omg07b02}~-~\ref{omg01b15}.

($2$) The approximated symmetry between $\varphi$ and $2\pi-\varphi$.

When the addition of two correlated CEPs is $2\pi$, i.e., $\varphi\leftrightarrow2\pi-\varphi$: At large (small) spatial scale, we observe the approximated mirror (coincidence) symmetry on the momentum spectrum in all chirping cases in both of high and low frequency fields. One can see the solid black (dashed blue and dashed-dotted red) lines for four set of correlated CEPs as $\pi/8\leftrightarrow15\pi/8$, $3\pi/8\leftrightarrow13\pi/8$, $5\pi/8\leftrightarrow11\pi/8$, $7\pi/8\leftrightarrow9\pi/8$ in Fig.~\ref{omg07mirror}, and also see three set of correlated CEPs as $2\pi/8\leftrightarrow14\pi/8$, $4\pi/8\leftrightarrow12\pi/8$, $6\pi/8\leftrightarrow10\pi/8$ in Fig.~\ref{omg01mirror}.
\begin{figure}[H]
\begin{center}
\includegraphics[width=\textwidth]{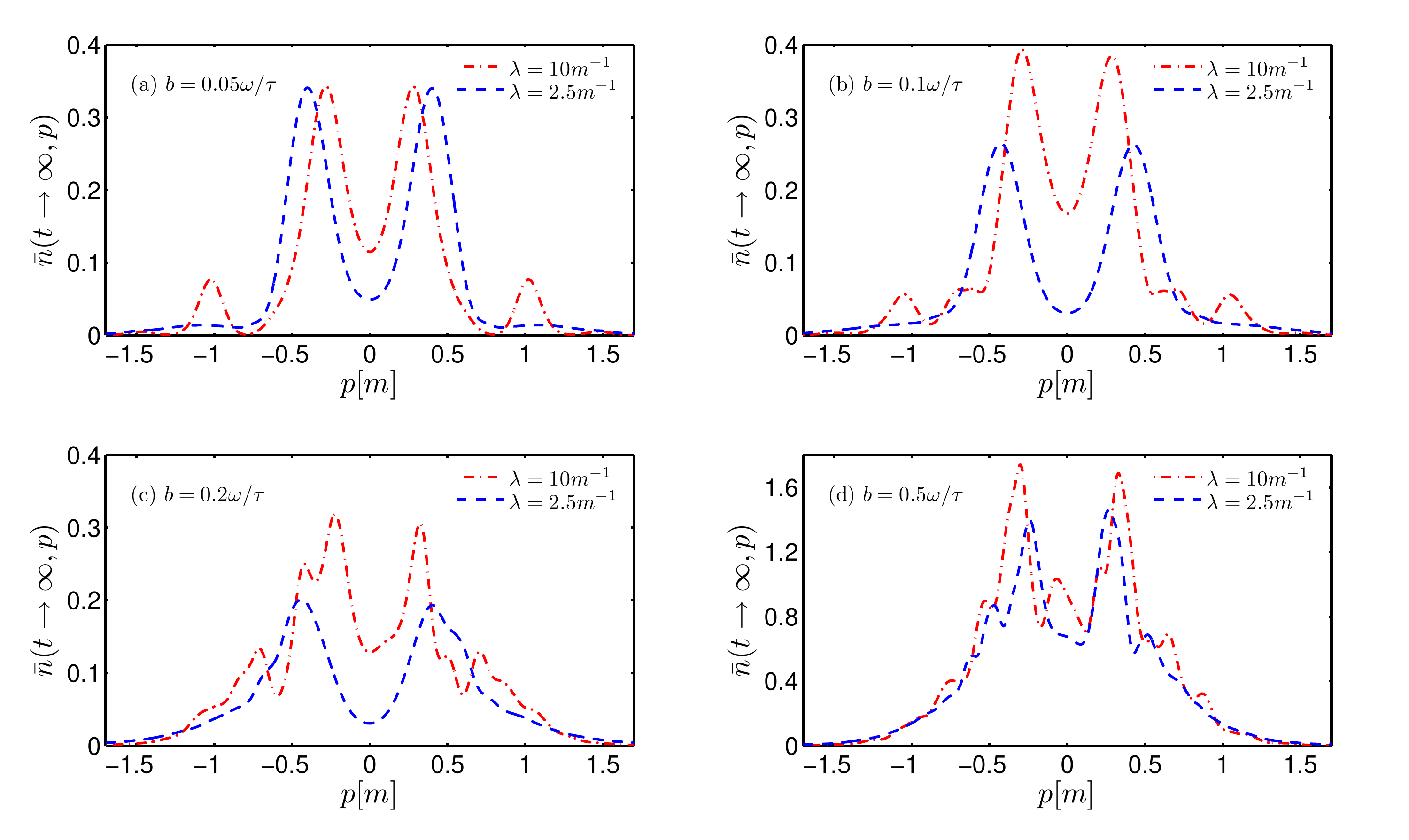}
\end{center}
\vspace{-12mm}
  \caption{(color online). The reduced momentum spectrum for various chirp parameters at different small spatial scales in the high frequency field when $\varphi=4\pi/8$. The chirp values are $b=0.05\omega/\tau=0.00078m^2$, $b=0.1\omega/\tau=0.0016m^2$, $b=0.2\omega/\tau \approx 0.0031 m^2$, and $b=0.5\omega/\tau \approx 0.0078 m^2$. Other field parameters are the same as in Fig.~\ref{omg07b02}.}
  \label{momentum07}
\end{figure}
At small spatial scale, another interesting symmetry occurs for individual CEP in both high and low frequency fields. We find that the individual momentum spectrum is approximated self symmetric with respect to $p=0$ for phase of either $4\pi/8$ or $12\pi/8$. This symmetry is caused by the combination of two different symmetries, the exact mirror $\varphi\leftrightarrow\pi+\varphi$ and approximated coincidence $\varphi\leftrightarrow2\pi-\varphi$ symmetry. In particular, this symmetry is more obvious in high frequency field, see Fig.~\ref{momentum07} and Fig.~\ref{momentum01}.

It is noted that these two types of approximated symmetries are more obvious in the high frequency field by comparing with that in the low frequency field. Meanwhile, they pronounce for without/small chirping cases and broken by strong oscillation on the momentum spectrum for large chirping cases.
\begin{figure}[H]
\begin{center}
\includegraphics[width=\textwidth]{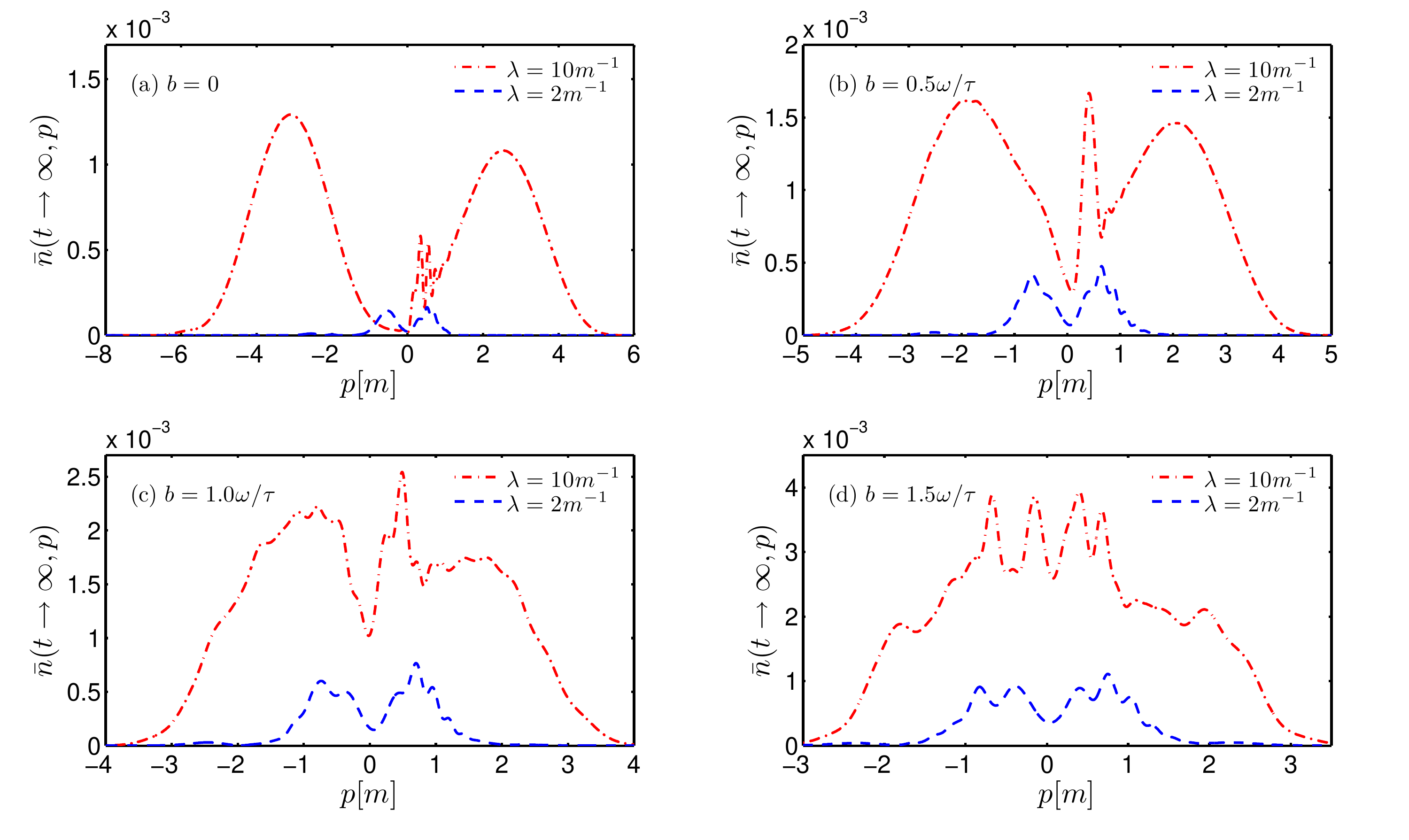}
\end{center}
\vspace{-12mm}
  \caption{(color online). The reduced momentum spectrum for various chirp parameters at different small spatial scales in the low frequency field when $\varphi=4\pi/8$. The chirp values are $b=0$, $b=0.5\omega/\tau=0.002m^2$, $b=1.0\omega/\tau=0.004m^2$ and $b=1.5\omega/\tau=0.006m^2$. Other field parameters are the same as in Fig.~\ref{omg01b0}.}
  \label{momentum01}
\end{figure}
No matter what kind of approximated symmetry, we can explain them theoretically by an approximation of the external field and the corresponding pseudodifferential operator. At large spatial scale, the electric field can be seen as only time dependent field approximately and the pseudodifferential operator can be rewritten as in Eq.~(\ref{pseudoDifflarge}). At small spatial scale, the pseudodifferential operator is mainly determined by the first term of Eq.~(\ref{pseudoDiff}) approximately because of the electric field exists in a very small space. Then we check the signs of all physical quantities according to the symmetry type which we observed on the momentum spectrum both at large and small spatial scales. The results are summarized in the Table~\ref{Table 3}. Finally we verify the validity of these different types of symmetries by substituting all physical quantities to the equations of motion in Eqs.~\eqref{pde:1}~-~\eqref{pde:4}, see the Appendix~\ref{Appendix} for more details.
\section{The reduced particle number density dependence on CEP}\label{result3}

To better explain that the sensitivity of momentum spectrum on CEP is significantly affected by symmetrical chirp and spatial scales, we examine the ratio of reduced particle number density dependence on CEP at fixed $p=0$ in both high and low frequency fields.

Here the ratio of reduced particle number density is defined via dividing the individual reduced particle number density $\bar{n}_{k}\left( p=0, t, \varphi_{k}\right)$ with the mean value of reduced particle number density $\bar{n}_{mean}\left( p=0, t, \varphi\right)$, and written as:
\begin{equation}\label{densityratio}
\bar{n}_{ratio}\left( p=0, t, \varphi\right) = \frac{\bar{n}_{k}\left( p=0, t, \varphi_{k}\right)}{\bar{n}_{mean}\left( p=0, t, \varphi\right)},
\end{equation}
where $\bar{n}_{mean}\left( p=0, t, \varphi\right) = \frac{\sum\limits_{k=0}^{15}{\bar{n}_{k}\left( p=0, t, \varphi_{k}\right)}}{16}$.

\begin{figure}[H]
\begin{center}
\includegraphics[width=\textwidth]{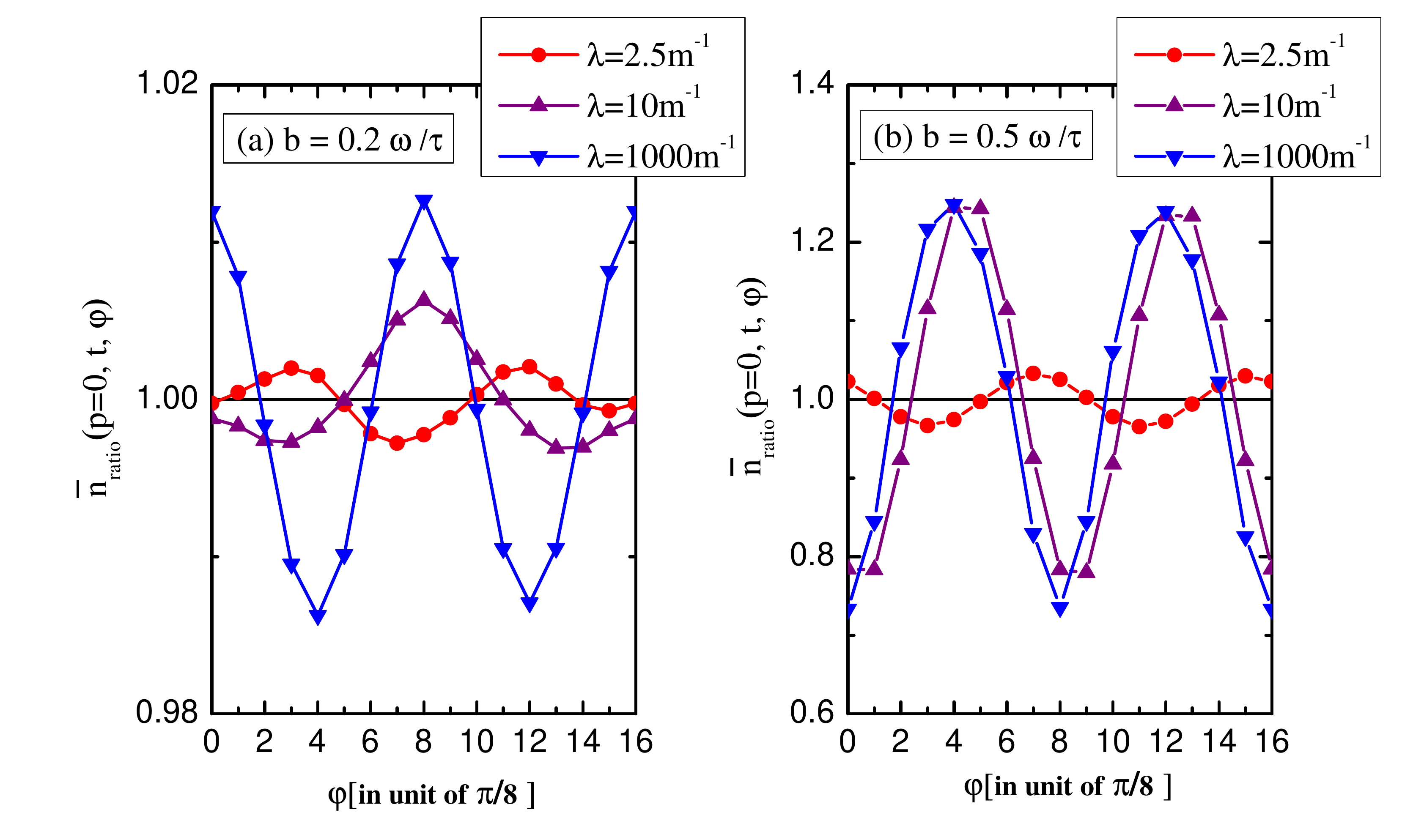}
\end{center}
\vspace{-13mm}
  \caption{(color online). The ratio of reduced particle number density dependence on CEP for fixed momentum value $p=0$ at different spatial scales for various chirp values in high frequency field. The chirp values are $b=0.2\omega/\tau \approx 0.0031 m^2$, $b=0.5\omega/\tau \approx 0.0078 m^2$, respectively. Other field parameters are the same as in Fig.~\ref{omg07b02}.}
  \label{momentumomg07}
\end{figure}

In high frequency field, at different spatial scales, the ratio of reduced particle number density depending on CEP for various chirp when $p=0$ are shown in Fig.~\ref{momentumomg07}.

We can see that the ratio for the largest chirping is much more greater than that for small chirping. It demonstrates that momentum spectrum is more sensitive to CEP in the largest symmetrical chirping case. In the point of view of spatial scale, the sensitivity of momentum spectrum on CEP is also suppressed by finite spatial inhomogeneity, and it reduces at small spatial scales, see dotted red and triangle purple lines in Fig.~\ref{momentumomg07}. However, the sensitivity is stronger even at small spatial scales for largest chirping by comparing with that for small chirping case in Fig.~\ref{momentumomg07}(a). These results are all consistent with the results of momentum spectrum which we obtained in Fig.~\ref{omg07b02} and Fig.~\ref{omg07b05}.

\begin{figure}[H]
\begin{center}
\includegraphics[width=\textwidth]{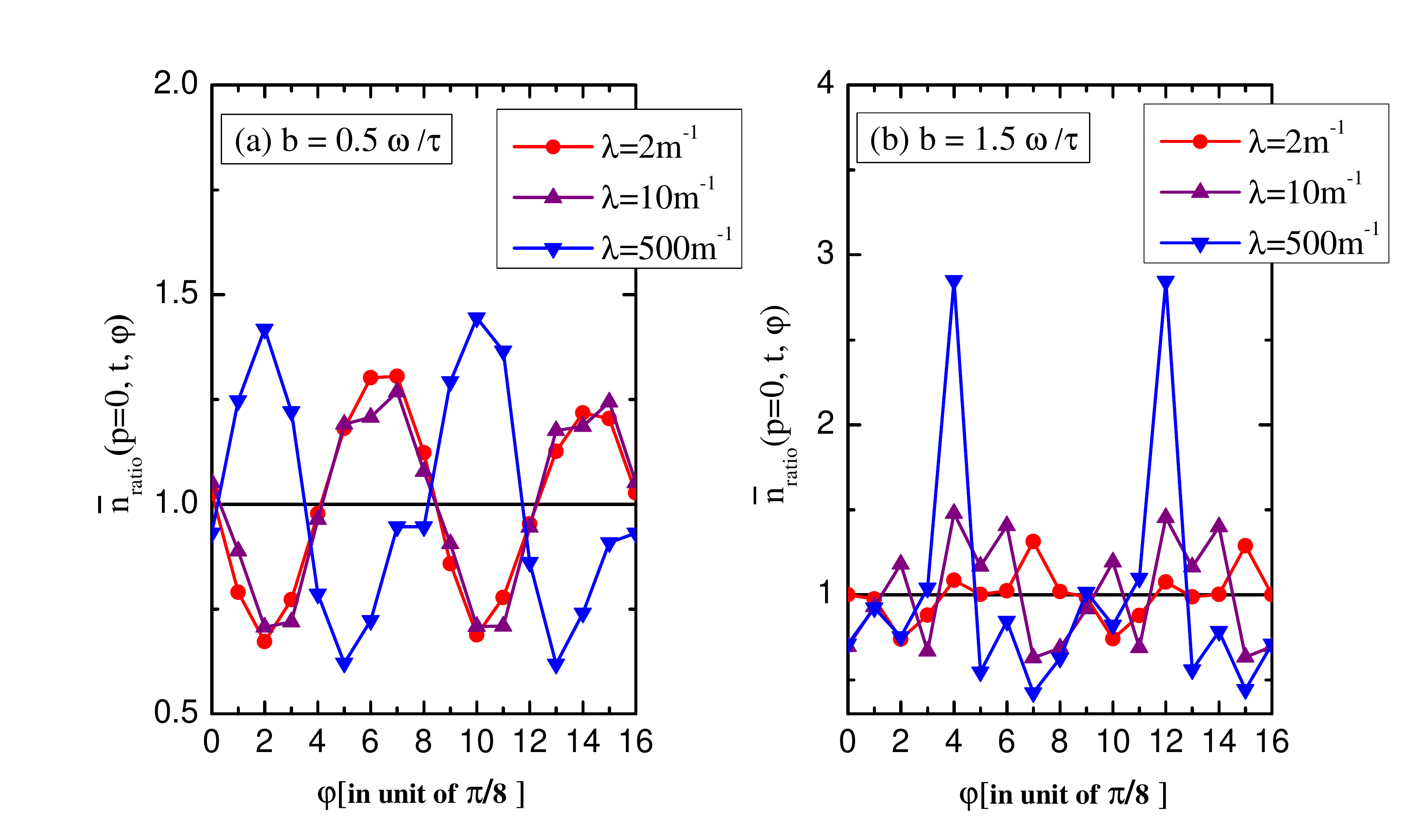}
\end{center}
\vspace{-13mm}
  \caption{(color online). The ratio of reduced particle number density dependence on CEP for fixed momentum $p=0$ at different spatial scales for various chirp values in low frequency field. The chirp values are $b=0.5\omega/\tau=0.002m^2$, $b=1.5\omega/\tau=0.006m^2$. Other field parameters are the same as in Fig.~\ref{omg01b0}.}
  \label{momentumomg01}
\end{figure}

In low frequency field, we also plot in Fig.~\ref{momentumomg01}, that the ratio of the reduced particle number density depend on CEP when $p=0$ at different spatial scales in cases of the  smallest and largest chirping.
We can observe great fluctuation on the value of ratio in both cases. It is clear that the momentum spectrum is very sensitive to CEP in low frequency field even with the very small chirping, and this sensitivity is stronger for the largest chirping. Moreover, the sensitivity of momentum spectrum for CEP is also suppressed by finite spatial scales, and reduces at small spatial scales.

Moreover, by the way, we have checked that the behavior of the reduced particle number density dependence on CEP which found in Fig.~\ref{momentumomg07} and Fig.~\ref{momentumomg01} is similar under other momentum value.

We can conclude that momentum spectrum is more sensitive to CEP in low frequency field by comparing with the high frequency field. In both of high and low- frequency fields, the sensitivity is most significant for largest chirping, meanwhile also suppressed by small spatial scales. These results are all consistent with that of the momentum spectrum in the previous section.

\section{CEP~effect on the reduced particle number}\label{result4}

In this section, we study the CEP effect on the reduced particle number in both high and low- frequency inhomogeneous fields with symmetrical frequency chirping.

\subsection{High frequency field}\label{result41}
\begin{figure}[H]
\begin{center}
\includegraphics[width=12cm]{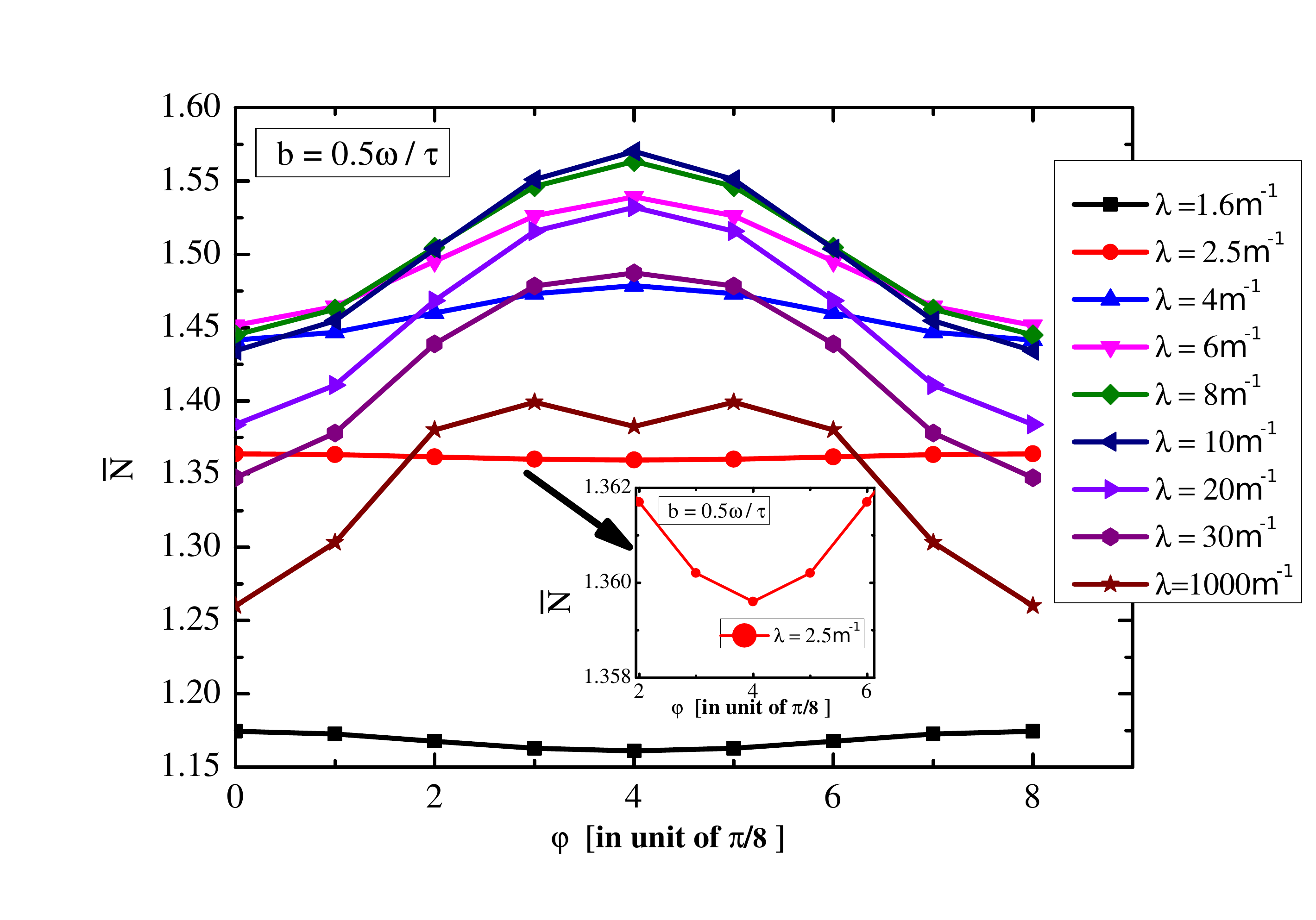}
\end{center}
\vspace{-15mm}
  \caption{(color online). The reduced particle number dependence on CEP at different spatial scales in high frequency field. Field parameters are the same as in Fig.~\ref{omg07b05}.}
  \label{omg07b05particle}
\end{figure}

When $b=0.2\omega/\tau \approx 0.0031 m^2$, the reduced particle number dependence on the CEP is very weak. Here we show the reduced particle number dependence on CEP at different spatial scales only when $b=0.5\omega/\tau \approx 0.0078 m^2$, see Fig.~\ref{omg07b05particle}.

We can see that at very small spatial scale $\lambda\leq 2.5m^{-1}$, the reduced particle number decreases at first and then increases gradually, minimum reduced particle number occurs for $\varphi=4\pi/8$. However, when $\lambda\textgreater 2.5m^{-1}$, it is enhanced significantly at the beginning until CEP reaches $\varphi=4\pi/8$, then it is decreased gradually for $5\pi/8\leqslant\varphi\leqslant\pi$. The maximum reduced particle number always occurs for $\varphi=4\pi/8$ except largest spatial scale $\lambda=1000m^{-1}$. It is just opposite to the changing patterns of cases of $\lambda\leq 2.5m^{-1}$. We have also noted that the largest reduced particle number can be obtained for the interval $6m^{-1}\leq\lambda\leq 10m^{-1}$. However, the maximal enhancement of reduced particle number is most pronounced for $\lambda=20m^{-1}$ when $0\leqslant\varphi\leqslant4\pi/8$.

Furthermore, we find that at all spatial scales the reduced particle number is exactly symmetric about $\varphi=4\pi/8$ in the interval $0\leqslant\varphi\leqslant\pi$. We would like to point out that this symmetry is induced by the approximated coincidence or/and mirror symmetry between $\varphi$ and $\pi-\varphi$ on the momentum spectrum. We have checked and found that it is also symmetric about $\varphi=\pi$ in the interval $0\leqslant\varphi\leqslant2\pi$. This symmetry can be understood by the exact mirror symmetry between $\varphi$ and $\pi+\varphi$ on the momentum spectrum.

\begin{figure}[ht]
\begin{center}
\includegraphics[width=12cm]{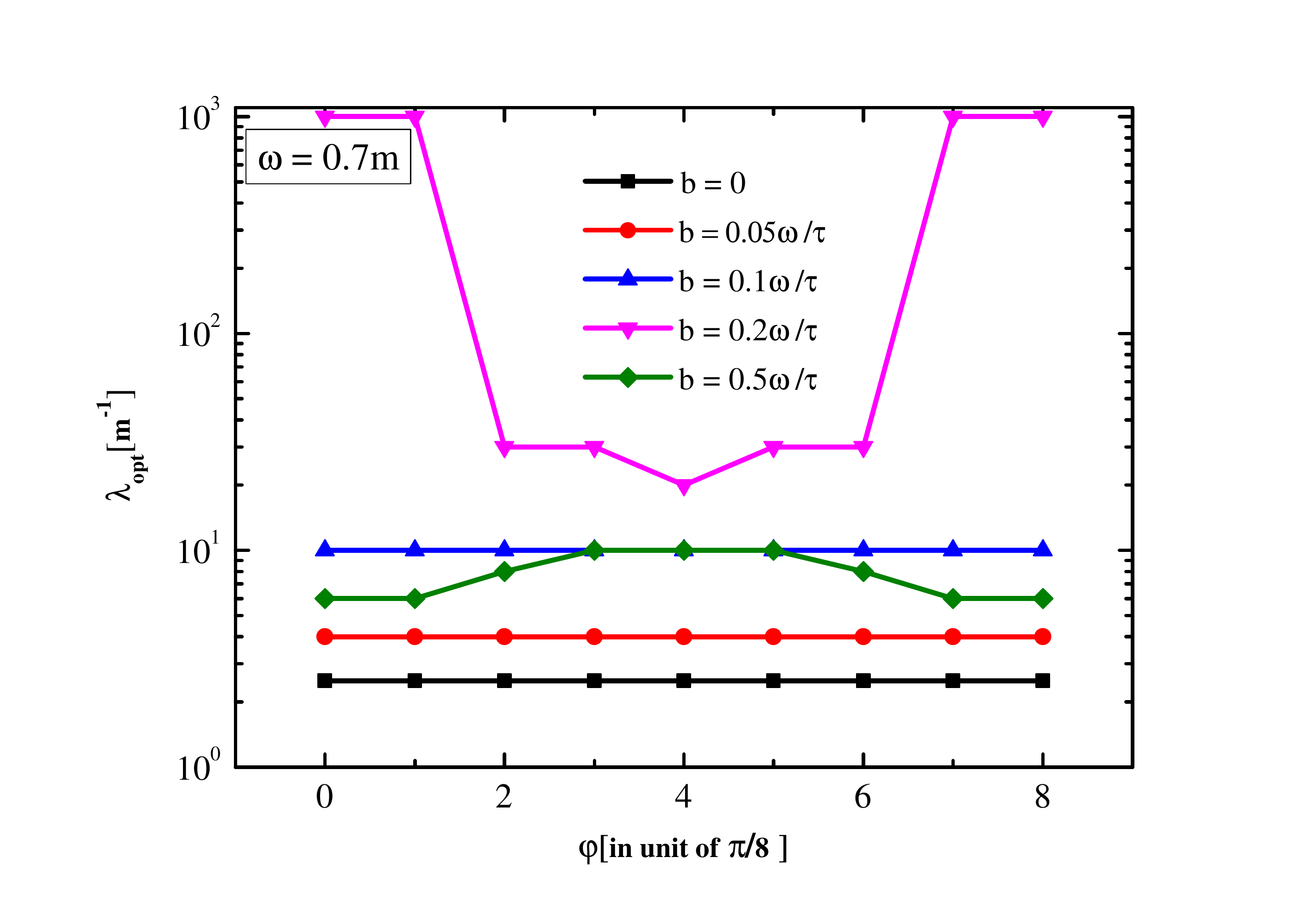}
\end{center}
\vspace{-15mm}
\caption{The optimal spatial scale of reduced particle number for various CEP with different chirp values: $b=0.05\omega/\tau \approx 0.00078 m^2$, $b=0.1\omega/\tau \approx 0.0016 m^2$, $b=0.2\omega/\tau \approx 0.0031 m^2$ and $b=0.5\omega/\tau \approx 0.0078 m^2$. Other field parameters are the same as in Fig.~\ref{omg07b05}.}
\label{omg07optimal}
\end{figure}

In Fig.~\ref{omg07optimal}, the optimal spatial scales of reduced particle number for various CEP with different chirping are plotted.
We find that optimal spatial scale is fixed for increasing CEP when $0\leq b\leq 0.1\omega/\tau \approx 0.0016 m^2$, see quadrangle black, dotted red and triangle blue lines in Fig.~\ref{omg07optimal}. However, it starts to sensitive for increasing CEP when chirp reaches $b=0.2\omega/\tau \approx 0.0031 m^2$. It shows again that, for large chirp $b=0.5\omega/\tau \approx 0.0078 m^2$, the optimal spatial scales are in the interval $6m^{-1}\leq\lambda_{opt}\leq 10m^{-1}$. These results indicate that the reduced particle number dependence on CEP is also determined by the symmetrical chirp value. Therefore, the combined effect of CEP and chirp is very crucial for pair production process.

\subsection{Low frequency field}\label{result22}
\begin{figure}[H]
\begin{center}
\includegraphics[width=\textwidth]{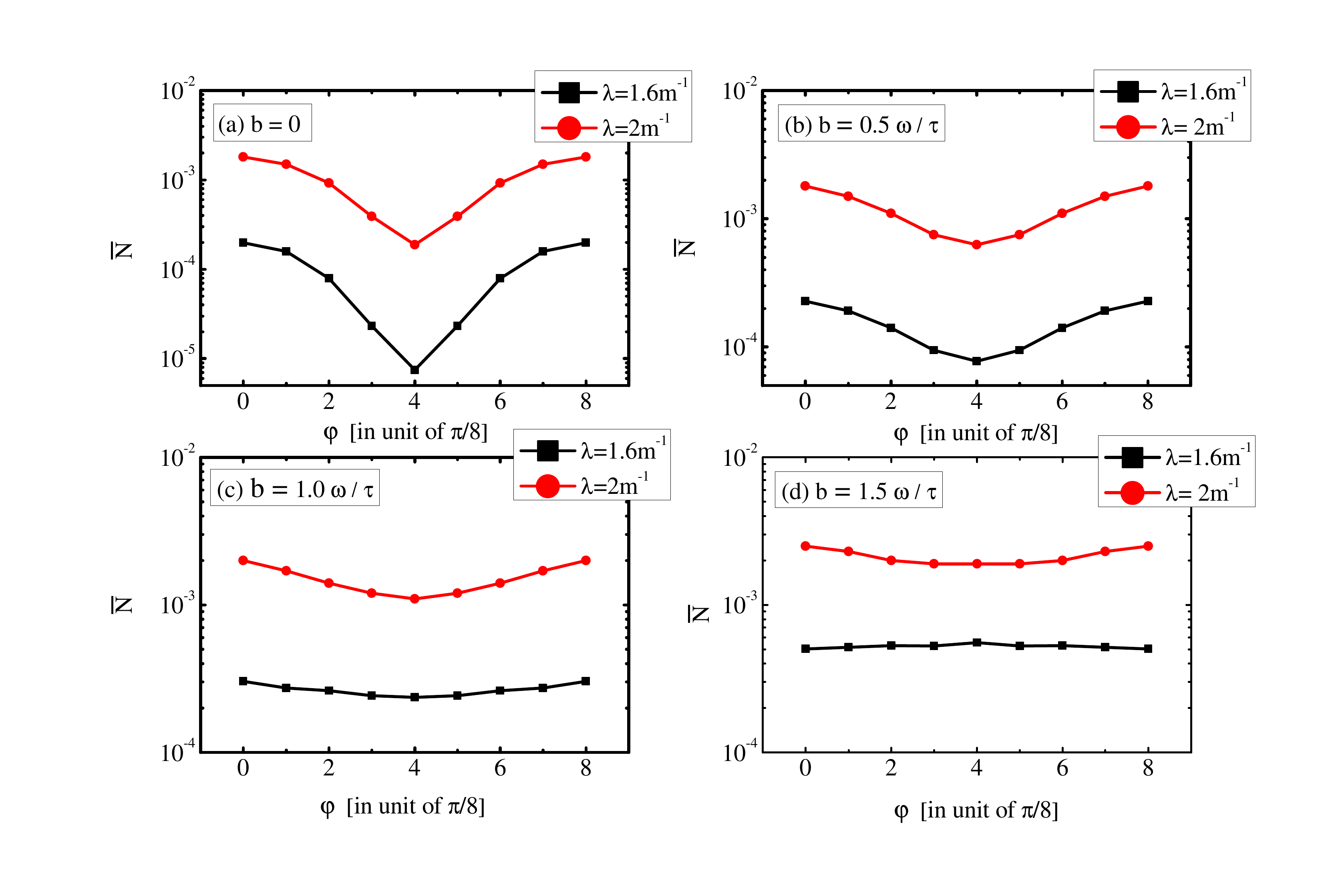}
\end{center}
\vspace{-1.5cm}
  \caption{(color online). The reduced particle number dependence on CEP for various chirp values at small spatial scales in low frequency field . The chirp values are $b=0$, $b=0.5\omega/\tau=0.002m^2$, $b=1.0\omega/\tau=0.004m^2$ and $b=1.5\omega/\tau=0.006m^2$. Other field parameters are the same as in Fig.~\ref{omg01b0}.}
  \label{omg01particle}
\end{figure}

\begin{figure}[H]
\begin{center}
\includegraphics[width=\textwidth]{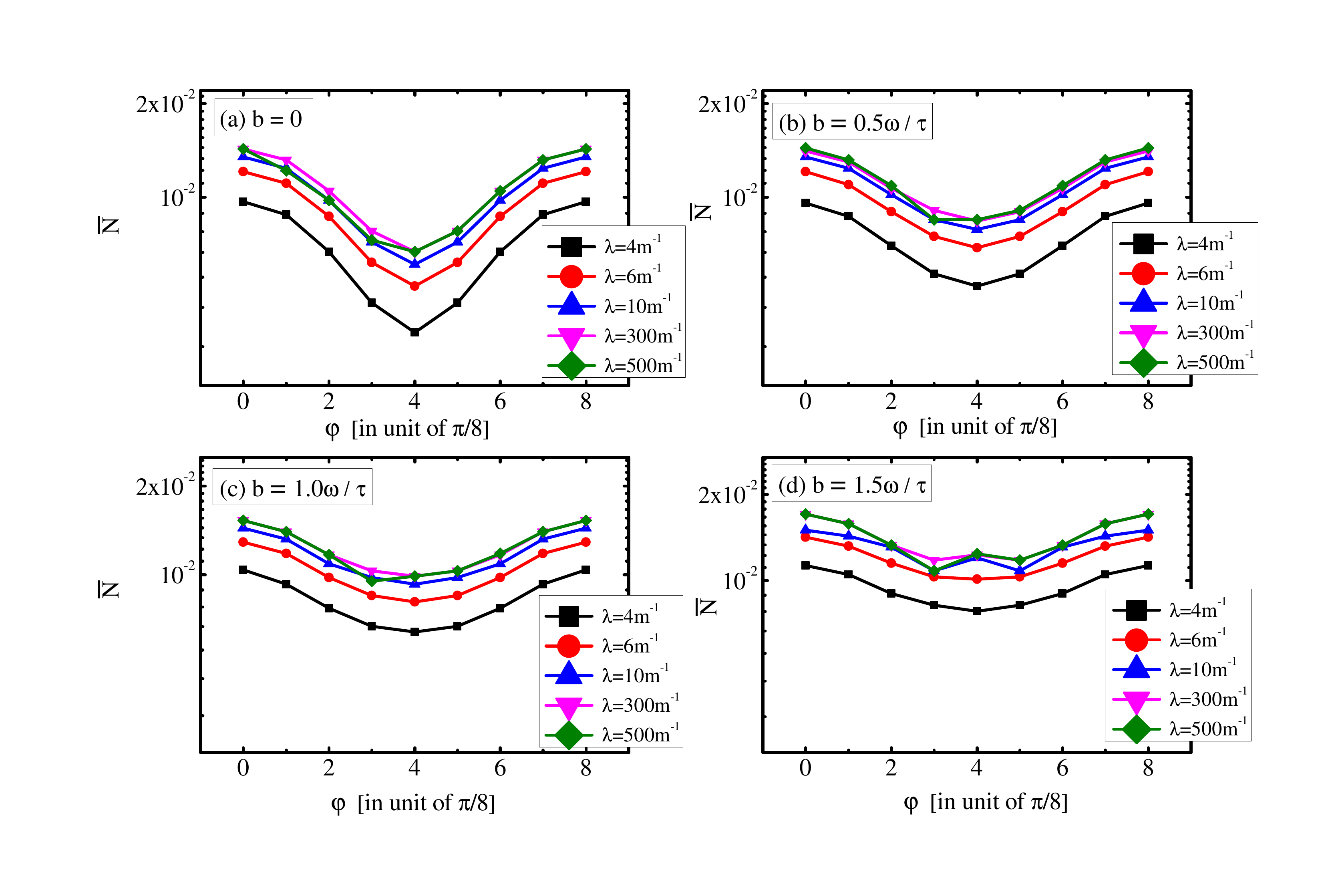}
\end{center}
\vspace{-1.5cm}
  \caption{(color online). The reduced particle number dependence on CEP for various chirp values at large spatial scales in low frequency field . The chirp values are $b=0$, $b=0.5\omega/\tau=0.002m^2$, $b=1.0\omega/\tau=0.004m^2$ and $b=1.5\omega/\tau=0.006m^2$. Other field parameters are the same as in Fig.~\ref{omg01b0}.}
  \label{omg01particle1}
\end{figure}

In Fig.~\ref{omg01particle} and Fig.~\ref{omg01particle1}, we also show the reduced particle number dependence on CEP for various chirping at small and large spatial scales, respectively. In both figures, the reduced particle number changes periodically in the interval $0\leqslant\varphi\leqslant\pi$. In particular, the reduced particle number decreases at the beginning and takes the minimum when $\varphi=4\pi/8$, then it increases again when $\varphi\geq5\pi/8$. Obviously, it is just opposite to the CEP dependence pattern of reduced particle in high frequency field, see Fig.~\ref{omg07b05particle}.
Furthermore, we can observe that the curves in the case of without chirping would be flat with increasing chirp at all spatial scales. It indicates that the reduced particle number is most sensitive to the increase of~CEP~in the case of without chirping, and this sensitivity decreases with increasing chirp. Because the reduced particle number is enhanced significantly with increasing chirp no matter whatever the CEP, therefore this sensitivity is depressed by larger chirping.
Now let us turn our attention to the enhancement of the reduced particle number in the interval $4\pi/8\leqslant\varphi\leqslant\pi$. We found that the enhancement with CEP is also different for large and small spatial scales. At small spatial scale, the reduced particle number is enhanced by $1$ to $1.5$ orders of magnitude with CEP for $b=0$ while it is only enhanced about $1.3$ times for the largest chirping, see Fig.~\ref{omg01particle}(a) and (d). At larger spatial scales, the reduced particle number is enhanced $2$ times for $b=0$ and this enhancement is decreased to $1.4$ times for the largest chirping, see Fig.~\ref{omg01particle1}(a) and (d). It is clear that the reduced particle number enhancement with increasing CEP is most obvious at small spatial scales without chirping. We can also conclude that the reduced particle number enhancement is more sensitive to CEP in low frequency field by comparing with the result of high frequency field.

Similarly, we find that the reduced particle number is symmetric about $\varphi=4\pi/8$ in the phase interval $0\leqslant\varphi\leqslant\pi$ except the largest spatial scale $\lambda=500m^{-1}$. This symmetry is caused by approximated symmetry on the momentum spectrum. Because three set of correlated CEPs as $\pi/8\leftrightarrow7\pi/8$, $2\pi/8\leftrightarrow6\pi/8$, $3\pi/8\leftrightarrow5\pi/8$ satisfy the phase relation $\varphi\leftrightarrow\pi-\varphi$, and cause approximated symmetry on the momentum spectrum. In this way, the reduced particle is also symmetric about $\varphi=4\pi/8$ correspondingly. However, this type of approximated symmetry on the momentum spectrum is not as strict as the exact mirror symmetry, therefore, it allows that this symmetry is also not strict on the reduced particle number consequently. As a result, we find the symmetry of reduced particle number with respect to $\varphi=4\pi/8$ does not hold at the largest spatial scale $\lambda=500m^{-1}$. Moreover, we examine and find that the reduced particle number is also exactly symmetric about $\varphi=\pi$ in the interval $0\leqslant\varphi\leqslant2\pi$ at all spatial scales. This symmetry is caused by the exact mirror symmetry on the momentum spectrum.

These results indicate that we may realize enhancement of pair production if we find out optimal values of CEP and chirping, which is crucial to optimize the momentum spectrum and the reduced particle number. Therefore, we have also checked out optimal values of CEP and chirping for the maximum reduced particle number in both high and low- frequency fields and discussed qualitatively. We will present and discuss our results in the next section.

\section{Discussions}\label{discussion}

In this section we would focus our discussions on the two features, i.e., two peaks structure on the momentum spectrum and the optimal values of CEP and chirping in detail.

\subsection{Two peaks structure on momentum spectrum}

\begin{figure}[H]
\begin{center}
\includegraphics[width=\textwidth]{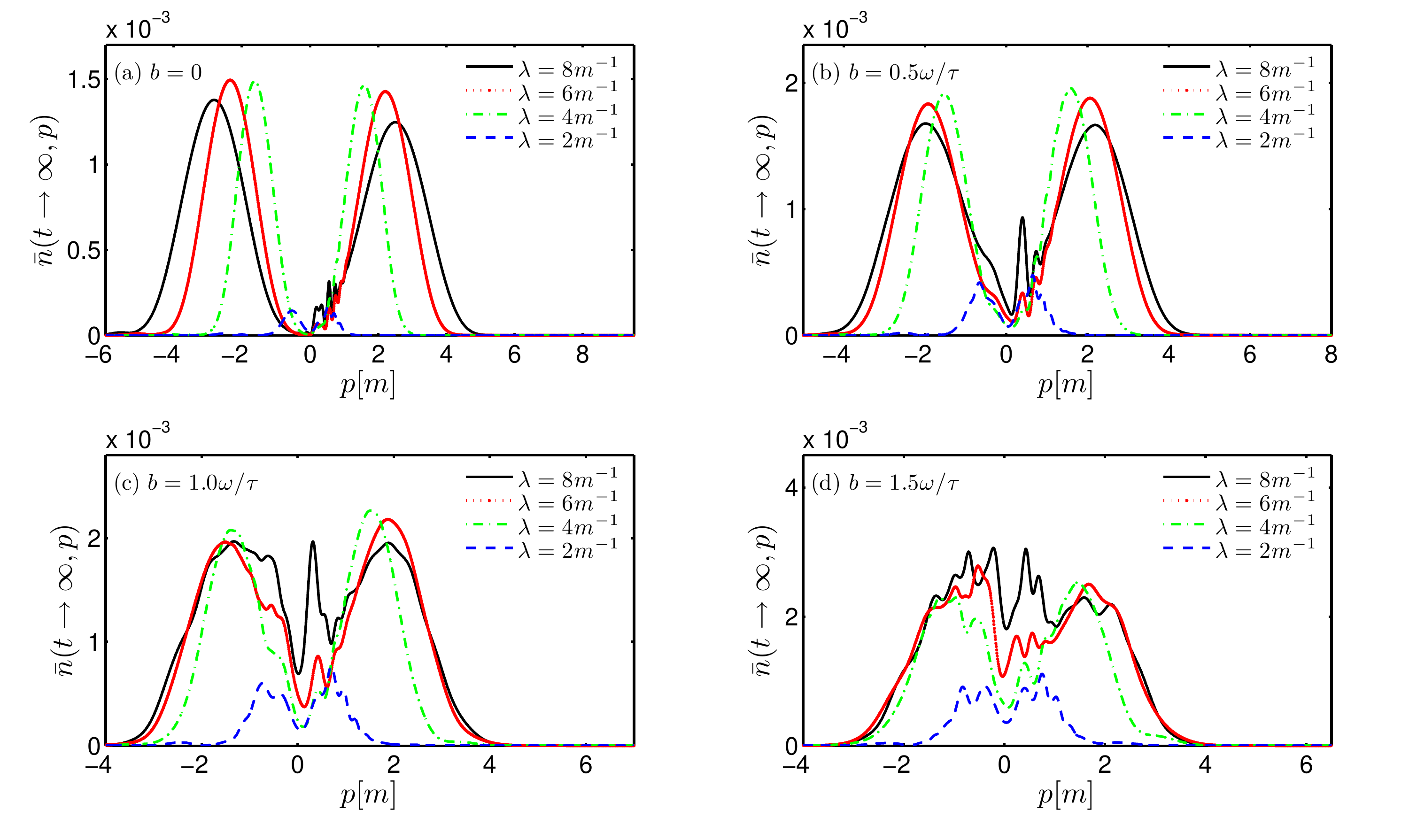}
\end{center}
\vspace{-12mm}
  \caption{(color online). The reduced momentum spectrum for various chirp values at small spatial scales in the low frequency field when $\varphi=4\pi/8$. The chirp values are $b=0$, $b=0.5\omega/\tau=0.002m^2$, $b=1.0\omega/\tau=0.004m^2$ and $b=1.5\omega/\tau=0.006m^2$. Other field parameters are the same as in Fig.~\ref{omg01b0}.}
  \label{omg01momentum}
\end{figure}

In both of high and low frequency fields, at small spatial scales, we have observed two momentum peaks on the momentum spectrum.
In the high frequency field, it is caused by the ponderomotive force that have been discussed extensively in Refs.~\cite{Kohlfurst:2017hbd,Ababekri:2020,
Mohamedsedik:2021}. Therefore, we just discuss the two momentum peaks which we observed in the low frequency field when $\varphi=4\pi/8$ here.
The reduced momentum spectrum for various chirp at small spatial scales are shown in Fig.~\ref{omg01momentum}.

When $b=0$, we obtain two momentum peaks at small spatial scales as shown in Fig.~\ref{omg01momentum}(a). The reason for these two peaks can be attributed to the changing of electric field shape. The oscillating mode of the electric field changes with CEP and turns into two opposite peaks form for $\varphi=4\pi/8$, see dashed blue line in Fig.~\ref{fieldomg01}(a). The particles are created from two opposite signed field peaks and leave field region in two opposite directions,  form the two opposite positioned peaks (left and right) on the momentum spectrum consequently. However, these peaks are vanished and replaced by strong oscillations for large chirp, see Fig.~\ref{omg01momentum}(c) and (d). This oscillations could be understood as the interference effect of created particles.
More importantly, when $b=0$, the two momentum peaks are far away more from the center with increasing spatial scales. It is just opposite to that in the high frequency field where the two momentum peaks are far away more from the center with decreasing spatial scales, one can see Fig.~1 (a) and (b) in Ref.~\cite{Mohamedsedik:2021}. Because in the low frequency field, more particles are created by electric field through tunneling when spatial scale is increased, and momentum spectrum also broadens correspondingly. However, it starts to shrink with increasing spatial scales when chirp is applied, see Fig.~\ref{omg01momentum}(b)~-~(d), which is similar to the phenomena of high frequency field we have mentioned above. It is probably because that low frequency field frequency contains more and more highfrequency components with increasing chirp, which are higher than its own original central frequency. In this sense, we have observed some similar features that found
on momentum spectrum in high frequency field.
\subsection{Optimal values of CEP and chirping}

\begin{figure}[H]
\begin{center}
\includegraphics[width=\textwidth]{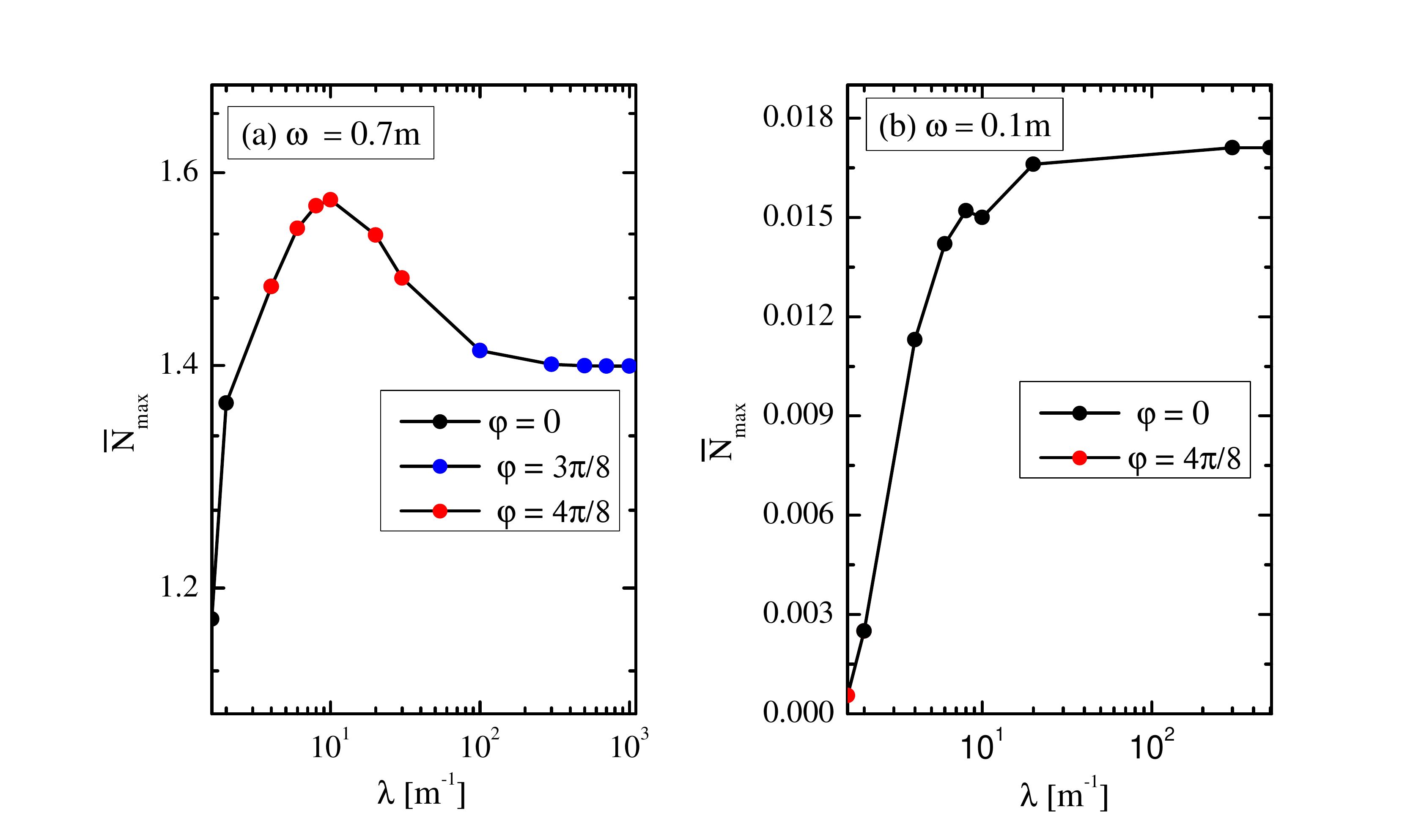}
\end{center}
\vspace{-12mm}
  \caption{(color online). The maximum reduced particle number dependence on CEP for high(a) and low(b) frequency electric fields at different spatial scales. The field parameters are $\omega=0.7m$, $\tau=45m^{-1}$, $\epsilon=0.5$, $b=0.5\omega/\tau=0.0078m^2$ for (a) and $\omega=0.1m$, $\tau=25m^{-1}$, $\epsilon=0.5$, $b=1.5\omega/\tau=0.006m^2$ for (b), respectively.}
  \label{maximum}
\end{figure}

In Fig.~\ref{maximum}, we also show optimal values of CEP and chirping for maximum reduced particle number in both of high and low- frequency electric fields. It is worthy to note that the reduced particle number is symmetric about $\varphi=4\pi/8$ in both of high and low- frequency fields as we mentioned in the previous section, and we have $\bar{N}_{\varphi=0}=\bar{N}_{\varphi=\pi}$ and $\bar{N}_{\varphi=3\pi/8}=\bar{N}_{\varphi=5\pi/8}$. In the legend of~Fig.~\ref{maximum} we have given the CEP in first quadrant for simplicity. We can see that the maximum reduced particle number is always occurring for the largest chirping at any spatial scale, however, it is different for CEP at different spatial scales in both of high and low- frequency fields.
In high frequency field, when $\lambda\leqslant2.5{m}^{-1}$, the reduced particle number decreases first and increases gradually with increasing CEP which we have mentioned in Fig.~\ref{omg07b05particle}. Therefore the maximum reduced particle number occurs for $\varphi=0$ or $\pi$. When $4{m}^{-1}\leqslant\lambda\leqslant30{m}^{-1}$, the reduced particle number increases until $\varphi=4\pi/8$ with CEP, the maximum obtains for $\varphi=4\pi/8$ correspondingly. At very large spatial scale, e.g.,
$\lambda=1000{m}^{-1}$, the maximum reduced particle number occurs for $\varphi=3\pi/8$ or $5\pi/8$. Nevertheless, it is worthy to emphasize that although the maximum reduced particle number obtaines for various CEP, however, in high frequency field, at fixed spatial scale, it is a little bit larger than the other particle number. For example, when $\lambda=4{m}^{-1}$, maximum particle number occurs for $\varphi=4\pi/8$ with $\bar{N}_{{max}{(\varphi=4\pi/8)}}=1.4787$, and the other non maximum particle numbers that corresponding to other CEP values are  $\bar{N}_{\varphi=3\pi/8}=\bar{N}_{\varphi=5\pi/8}=1.4732$, $\bar{N}_{\varphi=2\pi/8}=\bar{N}_{\varphi=6\pi/8}=1.46$.
We can conclude again that the reduced particle number has less dependence on CEP in high frequency field. This result can also be seen in the Fig.~4 of Ref.~\cite{Mohamedsedik:2021}.

\begin{figure}[H]
\begin{center}
\includegraphics[width=\textwidth]{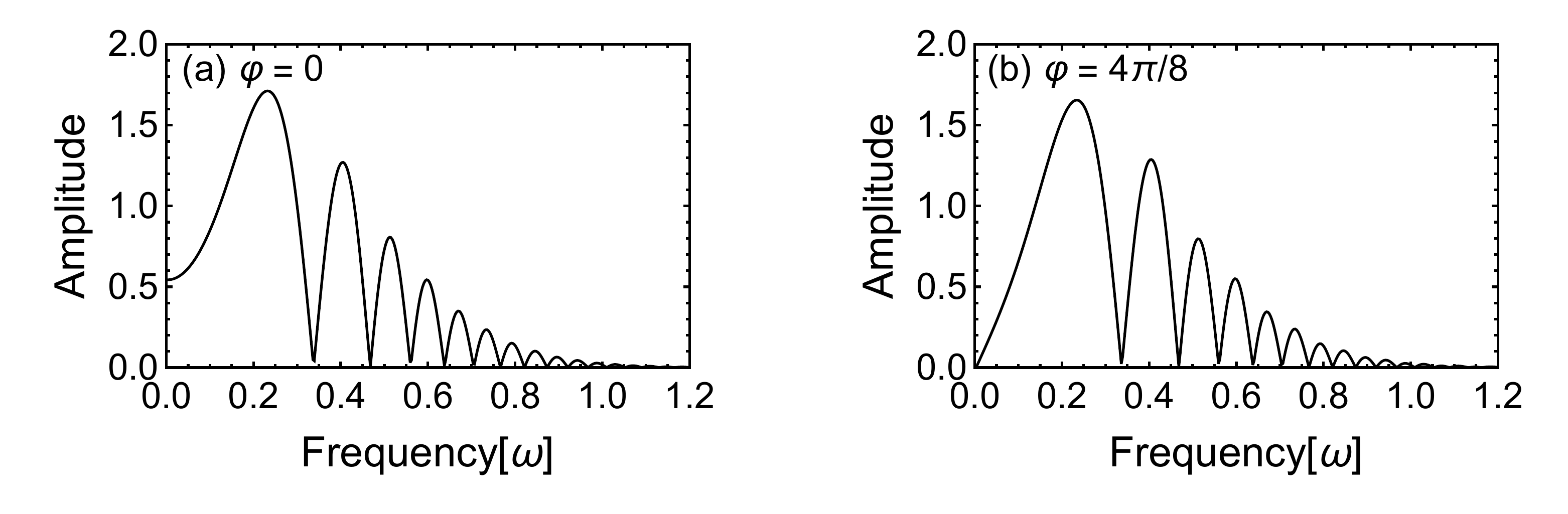}
\end{center}
\vspace{-12mm}
  \caption{Frequency spectrum of the electric field with the symmetrical chirp value $b=1.5\omega/\tau=0.006m^2$ for different two CEP values in the low frequency field. The other field parameters are $\epsilon=0.5$, $\omega=0.1m$, $\tau=25m^{-1}$.}
  \label{fourier}
\end{figure}

In the low frequency field, the maximum reduced particle number is always occurring for $\varphi=0$ or $\pi$ except the case of $\lambda=1.6{m}^{-1}$ where it is for $\varphi=4\pi/8$. It can also be seen in Fig.~\ref{omg01particle}(d). In fact, we can understand it by the frequency spectrum of electric field with symmetrical frequency chirp $b=1.5\omega/\tau=0.006m^2$ for different two CEPs of $\varphi=0$ and $\varphi=4\pi/8$ as shown in Fig.~\ref{fourier}.

We can see that when $\omega\leq0.35m$, the frequency amplitude starts from $0.55$ and increases to $1.7$ for $\varphi=0$ while it starts from $0$ and increases to $1.65$ for $\varphi=4\pi/8$. It is clear that for fixed frequency range there are more large amplitude frequency components when $\varphi=0$ than the case of $\varphi=4\pi/8$ one. In this way, the dynamical assisted role of symmetrical frequency chirp in the low frequency field which we mentioned in Ref.~\cite{Mohamedsedik:2021} is more significant in comparison with the case of $\varphi=4\pi/8$, and it can produce more particles in the field region. Consequently, we can observe that the maximum particle number in the case of $\varphi=0$ are higher than that in the case of $\varphi=4\pi/8$.

\section{Summary}\label{summary}

We have studied CEP effect on the pair production in spatial inhomogeneity electric field with frequency chirping by using the DHW formalism. We have exhibited the different impacts of the field parameters, such as CEP, frequency chirping and spatial scale, on the momentum spectrum and the reduced particle number. The main results are summarized as follows.

In high frequency field, with increasing CEP, the periodical changes are found about the interference effect on the momentum spectrum, the momentum peaks and the reduced particle number. The interference effect can be understood as the consequence of combined effects of CEP and chirping. The enhancement of the reduced particle number are most significant in case of largest chirping. Moreover, it is found that the reduced particle number is exactly symmetric about $\varphi=4\pi/8$ and also about $\varphi=\pi$ in phase intervals $0\leqslant\varphi\leqslant\pi$ and $0\leqslant\varphi\leqslant2\pi$, respectively, at fixed spatial scales. We also find the different CEP dependence of the optimal spatial scale in different case of chirping, i.e., in the case of without/small chirping, it is almost independent of CEP, however, in the case of large chirping, it depends on CEP.

In low frequency field, CEP has crucial effect on the shrink or broaden of the momentum spectrum. When chirping is fixed, we observed the periodical strong/weak variation of the interference effect with increasing CEP. There are two opposite positioned momentum peaks (left and right) at $\varphi=4\pi/8$ and it is discussed qualitatively. The reduced particle number change is also periodical with increaing CEP. The enhancement of the reduced particle number in the interval $4\pi/8\leqslant\varphi\leqslant\pi$ is evident and its enhancement degree with CEP is different at large and small spatial scales. At small spatial scale, the reduced particle number is enhanced over one order in the small/without chirping case with CEP while it is only enhanced about fewer times for the large chirping. At larger spatial scales, the reduced particle number is enhanced fewer times in the small/without chirping case with CEP but it is decreasing a little evenly for the large chirping.
These results indicate that the reduced particle number is most sensitive to the~CEP~in the case of without chirping, and this sensitivity decreases with chirping. At the small spatial scale, the sensitivity of particle number to CEP is more suppressed by the larger chirping in comparison with that at the large spatial scale.

More importantly and interestingly, in both of high and low frequency field, we observed several different types of symmetries on the two-correlated momentum spectrum such as exactly mirror symmetry, approximated coincidence or/and mirror symmetry for the different translations of two-correlated CEPs. And a specific approximated self symmetry on the individual momentum spectrum is also found at phase of either $4\pi/8$ or $12\pi/8$, where it is just the combination of exact mirror and approximated coincidence symmetries. It is noted that the symmetry is different for large and small spatial scales. We examined these symmetries by field model and the equations of motion in DHW, and the validity of the theoretical analysis conforms with our numerical results.

We also studied the reduced particle number density dependence on CEP at fixed momentum. It is found that particle number density is more sensitive to CEP in the largest symmetrical chirping case and this sensitivity is also suppressed by the small spatial scales in both high and low frequency fields. It is also noted that the particle number density is more sensitive to CEP in low frequency field in comparison with the high frequency field. Moreover, the optimal values of CEP and chirping are also investigated in both frequency fields. We found that the maximum reduced particle number is always occurring for the largest chirping at any spatial scale, however, it is different for CEP at different spatial scales.

These results presented in the paper suggest that, beside the chirping, the CEP plays also an important role on the pair production in spatial inhomogeneous electric field, therefore, the exhibited periodicity and the abundant symmetry of the momentum spectrum and the reduced particle number are the combinational effects of field parameters. The results and conclusions have potential implication by an extension to the more realistic field model such as electromagnetic field with multidimensional spatial scales and also the practical application for the created positron spectrum through controlling the field phase.

\begin{acknowledgments}
\noindent
This work was supported by the National Natural Science Foundation of China (NSFC) under Grant No.\ 11875007, 11935008.
The computation was carried out at the HSCC of the Beijing Normal University.
\end{acknowledgments}

\appendix
\section{Symmetry of momentum spectrum}\label{Appendix}

In this appendix, we give a detailed description for various type of symmetry which appears in the obtained momentum spectrum shown in the Sec.~\ref{result1} and revealed Sec.~\ref{symmetry}.

\subsubsection{Exact mirror symmetry}

The exact mirror symmetry holds for two phases as $\varphi\leftrightarrow\pi+\varphi$. It can be explained by the following consideration. First we note that when $\varphi$ becomes $\pi+\varphi$, the original electric field form in Eq.~(\ref{FieldMode}) changes sign as
\begin{equation}\label{FieldModepi}
\begin{aligned}
E\left(x,t;\pi+\varphi \right)=\epsilon \, E_{cr} \exp \left(-\frac{x^{2}}{2 \lambda^{2}} \right ) \exp \left(-\frac{t^{2}}{2 \tau^{2}} \right ) \cos[b |t| t + \omega t + (\pi+\varphi)]\\=-\epsilon \, E_{cr} \exp \left(-\frac{x^{2}}{2 \lambda^{2}} \right ) \exp \left(-\frac{t^{2}}{2 \tau^{2}} \right ) \cos(b |t| t + \omega t + \varphi)\\=-E\left(x,t;\varphi \right).
\end{aligned}
\end{equation}
Thus it leads to the momentum $p_x$ changes sign while $t$ does not, which means the $x$ should change the sign correspondingly. Moreover, because $\Omega(p_{x})$ is an even function of $p_x$ so that from Eq.~(\ref{vacuum-initial}), we have $\mathbbm{s}^{v} \left( x , p_{x} , t \right)$ even and  $\mathbbm{v}_{1}^{v} \left( x , p_{x} , t \right)$ odd, and finally it concludes that the particle number density is an even function about $p_x$
\begin{equation}\label{number density mirror}
n \left( x , p_{x} , t \right)_{E\left(\varphi\right)} = n \left( -x , -p_{x} , t \right)_{E\left(\varphi+\pi\right)},
\end{equation}
since that
\begin{equation}\label{particle number density mirror}
\frac{m  \mathbbm{s}^{v} \left( x , p_{x} , t \right) + p_{x}  \mathbbm{v}_{1}^{v} \left( x , p_{x} , t \right)}{\Omega \left( p_{x} \right)} = \frac{m  \mathbbm{s}^{v} \left( -x , -p_{x} , t \right) +p_{x}  \mathbbm{v}_{1}^{v} \left(- x , -p_{x} , t \right)}{\Omega \left( -p_{x} \right)}.
\end{equation}
By remembering in mind that $D_t\left( \partial_{t}, E_{x} , \partial_{p_{x}} \right) = D_t\left( \partial_{t}, -E_{x} , -\partial_{p_{x}} \right)$, $\mathbbm{p}^{v} \left( x , p_{x} , t \right) = - \mathbbm{p}^{v} \left( -x , -p_{x} , t \right)$ and $\mathbbm{v}_{0}^{v} \left( x , p_{x} , t \right) = \mathbbm{v}_{0}^{v} \left( -x , -p_{x} , t \right)$, we can conclude that when two phases is correlated as $\varphi\leftrightarrow\pi+\varphi$, the Eqs.~\eqref{pde:1}~-~\eqref{pde:4} of all the Wigner components are invariant under the exchange of $p_x$ and $-p_x$, therefore, we call it as the mirror symmetry. The odd/even signs, i.e., $-/+$, of all involved physical quantities under this exact mirror symmetry are given in Table~\ref{Table 3}.

\subsubsection{Approximated symmetry}

For another two cases, $\varphi\leftrightarrow\pi-\varphi$ and $\varphi\leftrightarrow2\pi-\varphi$, we observed the approximated coincidence or/and mirror symmetry in both of high and low frequency fields. It is worthy to note that, interestingly, the symmetry type is different at large and small spatial scales.

First we would like to discuss the pseudodifferential operator in Eq.~(\ref{pseudoDiff}) and obtain the approximated expressions for the electric field at the large and small spatial scales, respectively.
It is well known that when the spatial scale is very large, we treat the electric field approximately as only time dependent field by neglecting the spatial inhomogeneity. Thus the pseudodifferential operator in Eq.~(\ref{pseudoDiff}) can be rewritten as
\begin{equation}\label{pseudoDifflarge}
 D_t \approx \partial_{t} + e \int_{-1/2}^{1/2} d \xi \,\,\, E_{x} \left( t \right) \partial_{p_{x}} .
\end{equation}
On the contrary, when the spatial scale is very small, it means that the electric field exists only in a very small space, so that the expression of pseudodifferential operator is mainly determined by the first term of Eq.~(\ref{pseudoDiff}). In this way, we can write it approximately as
$D_t \approx \partial_{t}$.

In the following, we will discuss the corresponding symmetry under two different correlated translations of CEP, i.e., $\varphi\leftrightarrow\pi-\varphi$ and $\varphi\leftrightarrow2\pi-\varphi$, in both of the large and small spatial scales.

($1$) When the addition of two correlated CEPs is $\pi$, i.e., $\varphi\leftrightarrow\pi-\varphi$, the electric field is the odd under this phase exchange transformation as well as the reversed time.

At the large spatial scale, i.e., quasihomogeneous case, the electric field is approximate space-independent so that the $p_x$ can keep the same due to both reversed change of $t$ and $x$. It leads to the approximated coincidence symmetry on the momentum spectrum in both of the high and low frequency fields. In fact in this case, from Eq.~(\ref{vacuum-initial}) we have $\mathbbm{s}^{v} \left( x , p_{x} , t \right) = \mathbbm{s}^{v} \left( -x , p_{x} , -t \right)$ and $\mathbbm{v}_{1}^{v} \left( x , p_{x} , t \right) = \mathbbm{v}_{1}^{v} \left( -x , p_{x} , -t \right)$.

At small spatial scales, electric field leads to $p_x$ changes sign under the reversed $t$ only so that we have observed approximated mirror symmetry on the momentum spectrum in both of high and low frequency fields. Similarly in this situation, from Eq.~(\ref{vacuum-initial}), we have an even $\mathbbm{s}^{v} \left( x , p_{x} , t \right) = \mathbbm{s}^{v} \left( x , -p_{x} , -t \right)$ and an odd $\mathbbm{v}_{1}^{v} \left( x , p_{x} , t \right) = -\mathbbm{v}_{1}^{v} \left( x , -p_{x} , -t \right)$ expressions.

We can examine further the odd/even signs of all other physical quantities and Wigner components which corresponding to different case at different spatial scales by using the approximated expression of pseudodifferential operator as well as the equations of motion in Eqs.~\eqref{pde:1}~-~\eqref{pde:4}. They are shown in Table~\ref{Table 3} by Cases~\textbf{\Rmnum{2}(a)} and \textbf{\Rmnum{2}(b1)}, respectively.

($2$) When the addition of two correlated CEPs is $2\pi$, i.e., $\varphi\leftrightarrow2\pi-\varphi$, the electric field is the even under this phase exchange transformation as well as the reversed time. Therefore, at large (small) spatial scale, we can observe the approximated mirror (coincidence) symmetry in both of high and low frequency fields due to the complete similar reasons about the $-/+$ of $p_x$ mentioned above.
The detailed description for the invariant equations of DHW is given in Table~\ref{Table 3}, see Cases~\textbf{\Rmnum{3}(a)} and \textbf{\Rmnum{3}(b2)}.

For two specific phases at either $4\pi/8$ or $12\pi/8$, there exists an approximated individual symmetry by itself, because that it is the combinational symmetry by both types of exact mirror symmetry in Case \textbf{\Rmnum{1}} due to $12\pi/8-4\pi/8=\pi$ and also approximated coincidence symmetry in Case \textbf{\Rmnum{3}(b2)} due to $12\pi/8+4\pi/8=2\pi$, see Case \textbf{\Rmnum{4}(b3)}.
\begin{table}[H]
\caption{Three mutual symmetry for three different CEP translations, Case $\textbf{\Rmnum{1}}$, $\textbf{\Rmnum{2}}$ and $\textbf{\Rmnum{3}}$, and one individual self symmetry for phase $4\pi/8$ and $12\pi/8$, Case $\textbf{\Rmnum{4}}$. The (a) and (b) denote the large and small spatial scale. The (b1), (b2) and (b3) denote three different symmetry types at small spatial scale, respectively. The $-/+$ sign denotes the odd/even property of various physical variables or/and quantities under the phase translation.}
\centering
\setlength{\tabcolsep}{4mm}
\begin{ruledtabular}
\begin{tabular}{cccc}
\textbf{Case}& \textbf{Physical~~quantity} & \textbf{Sign} & \textbf{Symmetry~~type}                  \\ \hline
\textbf{\Rmnum{1}}    & $(t, p_{x}, x, \Omega, E_{x})$        & ($+$, $-$, $-$, $+$, $-$) & \multirow{3}{*}{exact mirror}  \\
$\varphi\leftrightarrow\pi+\varphi$ & $(\partial_{t}, \partial_{p_{x}}, \partial_{x}, D_{t})$ &($+$, $-$, $-$, $+$)&                           \\
all~~spatial~~scales & $(\mathbbm{s}$, $\mathbbm{v}_{1}$, $\mathbbm{p}$, $\mathbbm{v}_{0})$ & ($+$, $-$, $-$, $+$)    &                           \\ \hline
\textbf{\Rmnum{2}(a)} & $(t, p_{x}, x, \Omega, E_{x})$ & ($-$, $+$, $-$, $+$, $-$)     & \multirow{3}{*}{approximated coincidence} \\
$\varphi\leftrightarrow\pi-\varphi$& $( \partial_{t}, \partial_{p_{x}}, \partial_{x}, D_{t})$ &( $-$, $+$, $-$, $-$)&                           \\
 large~~spatial~~scales& $(\mathbbm{s}$, $\mathbbm{v}_{1}$, $\mathbbm{p}$, $\mathbbm{v}_{0})$ & ($+$, $+$, $-$, $+$)&                           \\ \hline
\textbf{\Rmnum{2}(b1)} & $(t, p_{x}, x, \Omega, E_{x})$        & ( $-$, $-$, $+$, $+$, $-$)   & \multirow{3}{*}{approximated mirror}   \\
$\varphi\leftrightarrow\pi-\varphi$& $(\partial_{t}, \partial_{p_{x}}, \partial_{x}, D_{t})$ &($-$, $-$, $+$, $-$)&                           \\
small~~spatial~~scales&$(\mathbbm{s}$, $\mathbbm{v}_{1}$, $\mathbbm{p}$, $\mathbbm{v}_{0})$ & ($+$, $-$, $+$, $+$)&                           \\ \hline
\textbf{\Rmnum{3}(a)} & $(t, p_{x}, x, \Omega, E_{x})$        & ($-$, $-$, $+$, $+$, $+$)    & \multirow{3}{*}{approximated mirror}   \\
$\varphi\leftrightarrow2\pi-\varphi$ & $(\partial_{t}, \partial_{p_{x}}, \partial_{x}, D_{t})$ &($-$, $-$, $+$, $-$) &                           \\
large~~spatial~~scales& $(\mathbbm{s}$, $\mathbbm{v}_{1}$, $\mathbbm{p}$, $\mathbbm{v}_{0})$ & ($+$, $-$, $+$, $+$)&                           \\ \hline
\textbf{\Rmnum{3}(b2)} & $(t, p_{x}, x, \Omega, E_{x})$        & ($-$, $+$, $-$, $+$, $+$) & \multirow{3}{*}{approximated coincidence}   \\
$\varphi\leftrightarrow2\pi-\varphi$& $(\partial_{t}, \partial_{p_{x}}, \partial_{x}, D_{t})$ &($-$, $+$, $-$, $-$)&                           \\
small~~spatial~~scales& $(\mathbbm{s}$, $\mathbbm{v}_{1}$, $\mathbbm{p}$, $\mathbbm{v}_{0})$ & ($+$, $+$, $-$, +)&                           \\ \hline
\textbf{\Rmnum{4}(b3)}& \multicolumn{1}{c}{} &   &   \\
~~$4\pi/8$ ~~or/and~~ $12\pi/8$ &         &\textbf{\Rmnum{1}}+\textbf{\Rmnum{3}(b2)}  & approximated self symmetry \\
small~~spatial~~scales &     &   &   \\

\end{tabular}
\end{ruledtabular}
\label{Table 3}
\end{table}

In summary we have found that no matter what kind of exact or/and approximated symmetry among the two correlated CEPs and also two specific phases of either $4\pi/8$ or $12\pi/8$, the invariant form of DHW works in this way. And all the numerical results present in the paper support and conform with these symmetries.
\appendix

\end{document}